\documentclass[twoside,onecolumn,draft]{IEEEtran}

\usepackage{latexsym,fancyheadings,amsmath,amssymb,amsthm,amsbsy,graphicx,epsfig,graphics,epsf}
\usepackage{epstopdf}
\usepackage{float}
\usepackage[tight,footnotesize]{subfigure}

\newtheorem{theorem}{Theorem}
\newtheorem{lemma}{Lemma}
\newtheorem{defi}{Definition}

\begin{document}
%
\title{Variable-length Splittable Codes with Multiple Delimiters}
%
%
%
%

\author{Anatoly~V.~Anisimov,~\IEEEmembership{Member,~IEEE} and Igor~O.~Zavadskyi~
\thanks{Manuscript received.}
\IEEEcompsocitemizethanks{\IEEEcompsocthanksitem Anatoly Anisimov is with the Department
of Mathematical Informatics, Taras Shevchenko National University of Kyiv, Kyiv,
Ukraine. \protect\\
E-mail: \underline {ava@unicyb.kiev.ua}%
\IEEEcompsocthanksitem Igor Zavadskyi is with the Department
of Mathematical Informatics, Taras Shevchenko National University of Kyiv, Kyiv,
Ukraine. \protect\\
E-mail: \underline {ihorza@gmail.com}

Copyright (c). Personal use of this material is permitted.  However, permission to use this material for any other purposes must be obtained from the IEEE by sending a request to pubs-permissions@ieee.org.
}}

\IEEEcompsoctitleabstractindextext{%
\begin{abstract}
Variable-length splittable codes are derived from encoding  sequences  of ordered integer pairs, where one of the pair's components is upper bounded by some constant, and the other one is any positive integer. Each pair is encoded by the concatenation of two fixed independent prefix encoding functions applied to the corresponding components of a pair. The codeword of such a sequence of pairs consists of the sequential concatenation of corresponding pair's encodings. We call such codes splittable. We show that Fibonacci codes of higher orders and codes with multiple delimiters of the form $011\ldots10$ are splittable. Completeness and universality of multi-delimiter codes are proved. Encoding of integers by multi-delimiter codes is considered in detail. For these codes, a fast byte aligned decoding algorithm is constructed. The comparative compression performance of Fibonacci codes and different multi-delimiter codes is presented. By many useful properties, multi-delimiter codes are superior to Fibonacci codes.
\end{abstract}

}

\maketitle

\IEEEdisplaynotcompsoctitleabstractindextext

\IEEEpeerreviewmaketitle

\section{Introduction}
The present period of the information infrastructure development is distinguished by the active interaction of various computer applications with huge Information Retrieval Systems. This activity actualizes the demand for efficient data compression methods that on one hand provide satisfactory compression rate, and, on the other, support fast search operations in compressed data. Along with this, the need for code robustness in the sense of limiting possible error propagations has been also strengthened.

As is known, in large textual databases classical Huffman codes \cite {huffman}, when applied to words considered as symbols, show good compression efficiency approaching to the theoretically best. Unfortunately, Huffman's encoding does not allow a fast direct search in compressed data by a given compressed pattern. At the expense of losing some compression efficiency, this was amended by introducing byte aligned tagged Huffman codes.  They are Tagged Huffman Codes \cite{THufmann}, End-Tagged Dense Codes (ETDC) \cite{ETDC}, and (s,c)-Dense Codes (SCDC) \cite{SCDC}. In these constructions, codewords are represented as sequences of bytes, which along with encoded information incorporate flags for the end of a codeword.

The alternative approach for compression coding stems from using  Fibonacci numbers of higher orders. The mathematical study of Fibonacci codes was started in the pioneering paper \cite{AF}. The authors first introduced a family of Fibonacci codes of higher orders with the emphasis on their robustness. They proved completeness and universality of these codes.

The strongest argumentation for the use of  Fibonacci codes of higher orders in data compression is given in \cite{Kl}, \cite{Kl1}. For these codes, the authors developed fast byte aligned algorithms for decoding \cite{KlDec} and search in  compressed text \cite{KlSearch}. They also showed that Fibonacci codes have better compression efficiency comparing with ETDC and SCDC while still being somewhat inferior in decompression and search speed even if  byte aligned algorithms are applied.

Evidently, the structure of a code strongly depends on the form of initial data representation. Note that in their constructions many integer encodings use two-parted information. For instance, the simplest Run-Length Codes use pairs (\emph{the count of a symbol in a  run, symbol}). The famous Elias \cite{El}, Levenshtein \cite{Lev} and many other codes that use their own length \cite{Salomon} exploit the pairing integer information (\emph{bit length, binary representation}). The Golomb \cite{Golomb} and the Golomb-Rice \cite{Rice} codes use pairs (\emph{quotient, remainder}) under integer division by a fixed number.

So, we argue that many code constructions fit into the general scheme as follows:

(i) According to some mathematical principle, each element of the input alphabet is put into one-to-one correspondence with the sequence of ordered integer pairs. Some relationships inside pairs and among pairs could be specified.

(ii) For encoding pairs, some variable-length uniquely decodable function is chosen.

(iii) To obtain the resultant codeword of a sequence of pairs, the corresponding codewords of pairs are concatenated in direct or reverse order.

(iv)  A special delimiter could be appended to the obtained binary sequence.

This general scheme could be specified in many ways. One of such variants with the emphasis on splitting a code into simpler basic components is considered in this presentation.

We introduce and study a family of binary codes that are derived from encoding sequences of ordered integer pairs with restrictions on one of the pair's component. Namely, we consider the initial data representation of the form $(\Delta_{1}, k_{1})\ldots(\Delta_{t}, k_{t})$, where all integers $\Delta_{i}$  are upper bounded by some constant $d$,  values $k_{i}$ are not bounded, $0\leq\Delta_{i}\leq d$, $0<k_{i}$, $i=1,\ldots,t$. Each pair is encoded using the concatenation of two fixed independent prefix encoding functions applied to the corresponding components of a pair. A codeword consists of the sequential concatenation of those pair's encodings. We call such codes splittable. Depending on tasks to be solved, one can choose a variety of coding functions to encode each pair $(\triangle, k)$ . This way we construct a code, which we call a $(\triangle, k)$-code.

In the same way by using the dual representation $(k_{1},\triangle_{1}),\ldots,(k_{t},\triangle_{t})$, we define $(k,\Delta)$-codes.

The families of $(\Delta,k)$ and $(k,\Delta)$-codes constitute the set of splittable codes. Giving such a name to considered codes we want to stress that the structure of a  code reflects the splittable nature of the initial data representation by simpler integral parts. Splittable codes could be considered as a generalization of  Golomb's codes, which contain only one $(k,\Delta)$-pair.

Splittable codes are well structured. Each codeword, including delimiters, is the concatenation of an integral number  of corresponding $(\Delta,k)$   or $(k,\Delta)$-pairing encodings. This regularity of a code structure also facilitates proving its important properties, such as completeness, universality, and  density.

In spite of the fact that $(\Delta,k)$  and $(k,\Delta)$-sequences carry the same information about coded data, their encodings could be very different. We prove that any Fibonacci code belongs to the class of $(k,\Delta)$-codes and cannot be any $(\Delta,k)$-code.

An important family of $(\Delta,k)$-codes are variable length codes with multiple delimiters. These codes are the main subject of our study.

A delimiter is a synchronizing string that makes it possible to uniquely identify boundaries of codewords under their concatenation.  In our case, each delimiter consists of a run of consecutive ones surrounded with zero brackets. Thus, delimiters have the form $01\ldots10$. A delimiter either can be a proper suffix of a codeword, or it arises as the concatenation of the codeword ending zero and a codeword of the form $11\ldots10$.
 The number of ones in delimiters is defined by a given fixed set of positive integers $m_{i}, i=1,2,\ldots,t$.
The multi-delimiter code of that form is  denoted by $D_{m_{1},\ldots,m_{t}}$.  We prove that any multi-delimiter code $D_{m_{1},\ldots,m_{t}}$  is a $(\Delta, k)$- code and thus splittable.

By their properties, multi-delimiter codes are close to Fibonacci codes of higher orders.  We prove completeness and universality of those codes. There also exists a bijection between the set of natural numbers and any code $D_{m_{1},\ldots,m_{t}}$. This bijection is implemented by simple encoding and decoding procedures. For practical use, we present a byte aligned decoding algorithm, which has better computational characteristics than that of Fibonacci codes developed in \cite{Kl1}.

As shown in \cite{Kl1}, the Fibonacci code of order three, denoted by Fib3, is the most effective for the text compression.From our study it follows  that the simple code $D_{2}$ with one delimiter $0110$ has asymptotically higher density as against Fib3, although it is slightly inferior in compression rate for realistic alphabet sizes of natural language texts.

We also note that by varying delimiters for better compression we can adapt multi-delimiter codes
to a given probability distribution and an alphabet size.
Thus, for example, we compare the codes $D_{2,3}$, $D_{2,3,5}$ and $D_{2,4,5}$ with the code Fib3.
Those multi-delimiter codes are asymptotically less dense than Fib3.
Nevertheless, alphabet sizes of the texts used in practice are relatively small, from a few thousands up to a  few millions words. For texts of such sizes the mentioned above multi-delimiter codes outperform the Fib3 code in compression rate. The conducted computational experiment shows that, for example, the code $D_{2,3,5}$ gives the average codeword length by $2-3\%$ shorter than the Fib3 code when encoding the Bible and some other known texts. Even in encoding one of the largest up to date natural language text corpus of English Wikipedia, the code $D_{2,3,5}$  is still superior as well as the codes $D_{2,3}$ and $D_{2,4,5}$.

Multi-delimiter codes, like Fibonacci codes, are static codeword sets not depending on any probability distribution. For a multi-delimiter code there exists an easy procedure for generating all words of a given length. Therefore, these codes allow an easy vocabulary representation for compression and decompression procedures. To create the vocabulary, one only needs to sort symbols according to the probabilities of their occurrences.

Due to robust delimiters, multi-delimiter codes are synchronizable with synchronization delay at most one codeword.

Properties of multi delimiter codes mainly rely on a finite set of special suffixes. Sets of words with a given fixed suffix, which cannot occur in other places of a word, are known as pattern codes. Properties of these codes such as synchranizability, completeness, universality, the average codeword length have been intensively studied \cite{Lakshman}-\cite{Lima}. Multi-delimiter codes even with one delimiter are not pattern codes, although they belong to the class of universal codes that are regular languages \cite{Capoc}.

The structure of this presentation is as follows. Prior to the introduction of splittable codes, we precede with the consideration of two simpler codes of that type. In Section 3 with the purpose to show how $(\Delta,k)$-constructions arise in integer encodings,  we briefly consider a specific integer representation using the two-base numeration system with the main radix 2 and the auxiliary radix 3.
This representation yields a typical $(\Delta, k)$-code  with restrictions given by inequalities $0\leq\Delta\leq2$, $0<k$. This code is universal, but it is not complete.
In section 4 we show that it can be embedded into the larger one-delimiter code set $D_{2}$, which is complete.

In section 5 we introduce splittable codes, and discuss $(\Delta,k)$ versus  $(k,\Delta)$-codes. We argue that $(\Delta,k)$-codes have some advantages comparing with  $(k,\Delta)$-codes. That includes the possibility to form a wider variety of short codewords and more efficient codeword separation.

In section 6 we introduce multi-delimiter codes $D_{m_{1},\ldots,m_{t}}$. We prove the mentioned above main properties of these codes: being a $(\Delta, k)$-code, completeness, and universality.

A bijective correspondence between the set of natural numbers and the codewords of any code $D_{m_{1},\ldots,m_{t}}$   is established in the next section. For multi-delimiter codes we present simple algorithms for encoding integers and decoding codewords. With the purpose to accelerate the procedure of decoding we describe the general scheme of a byte aligned algorithm. Using the code $D_{2}$ as the representative of the considered family of codes a byte aligned decoding algorithm is presented in detail in Section 8.

Comparative density characteristics of different multi-delimiter codes and the code Fib3 are given in Section 9.

Our conclusion is the following. The introduced multi-delimiter codes form a rich adaptive family of robust data compression codes that could be useful in many practical applications.

\section{Definitions and notations}

By $\{0,1\}^*$ denote the set of all strings in the alphabet $\{0,1\}$.
Let $m$ be a non-negative integer. Denote by $1^{m}$  (respectfully  $0^{m}$) the sequence consisting of $m$ consecutive ones (respectfully $m$ zeros).

The empty string corresponds to the value $m = 0$.

A run of consecutive ones  in a word $w$ is called isolated if  it is a prefix of this word ending with zero, or it is its suffix starting with zero, or it is a substring of $w$ surrounded with zeros, or it coincides with $w$.

For a word $w\in\{0,1\}^{*}$ its length is denoted by $|w|$.

A code is a set of binary words.

A code is called prefix (prefix-free) if no codeword could be a prefix of another codeword.

A code is called uniquely decodable (UD) if any concatenation of codewords is unique. Each prefix code has UD property.

A code is called complete if its any extension leads to not UD code.

Let $(\Delta_{0},k_{0})...(\Delta_{t},k_{t})$ be a sequence of ordered integer pairs, where $0\leq\Delta_{i}\leq d, 0<k_{i}$. For simplicity, in the sequel, pairs $(\Delta_{i},k_{i})$ of that type are called $(\Delta,k)$-pairs, and a sequence of such pairs is called a $(\Delta,k)$-sequence. Symbols $\Delta$  and $k$ can be viewed as names of variables corresponding to values  $\Delta_{i}$  and $k_{i}$.

We encode values $\Delta$   and $k$ by some fixed prefix binary codes. The codeword of a $(\Delta,k)$-pair is the concatenation of codewords corresponding to parameters $\Delta$   and $k$. The codeword of a $(\Delta,k)$-pair is called the $(\Delta,k)$-group.

In analogous way by changing the order in pairs we define $(k,\Delta)$-pairs, $(k,\Delta)$-sequences, and $(k,\Delta)$-groups.

Fibonacci numbers of order $m\geq1$, denoted by $F^{(m)}_{i}$, are defined by the recurrence relation:

$F^{(m)}_{n}=F^{(m)}_{n-1}+F^{(m)}_{n-2}+...+F^{(m)}_{n-m}$ for $n>1$

$F^{(m)}_{1}=1, F^{(m)}_{n}=0$ for $-m<n\leq0$.

The Fibonacci code of order $m$, denoted by Fib\emph{m}, is the set consisting of the word $1^{m}$  and all other binary words that contain  exactly one occurrence of the substring $1^{m}$, and this occurrence is the word's suffix \cite{Kl1}.

For any $n$ the Fibonacci code Fib$m$ contains exactly $F^{(m)}_{n}$ codewords of the length $n+m$.

\section {Lower $(2,3)$-representation of numbers}

Representation of numbers in the mixed two-base numeration system using the main radix 2 and the auxiliary radix 3 was first introduced in \cite{Ani1}. Prefix encoding of integers using this representation was studied in \cite{Ani2}. The so-called lower (2,3)-representation of numbers, which is a modification of the general (2,3)-representation, was introduced in \cite{AZ}.  Let us briefly describe its essence.

Let ${\mathbb{N}_{2,3}}$ be the set of natural numbers that are coprime with 2 and 3,
 $x\in {\mathbb{N}_{2,3}}$,   $x>1$,  $n =\lfloor\log_{2}x\rfloor$, $1\leq m\leq n$.

A very simple idea stands behind the $(2,3)$-integer representation.  Note that for any whole positive number $m$ integers $2^{m}$ and $2^{m-1}$ give different residues modulo 3. Therefore, $x$ can be uniquely represented in one of the forms
 $2^{m} + 3^{k}x_{1}$  or  $2^{m-1} + 3^{k}x_{1}$, where $x_{1}$ also belongs to ${\mathbb{N}_{2,3}}$ and $k \geq 1$.

In the general $(2,3)$-representation of $x$ the maximal value is chosen for $m$, $m=\lfloor\log_{2}x\rfloor$. In the lower $(2,3)$-representation we use the shifted value, $m=\lfloor\log_{2}x\rfloor -1$. Such a choice for $m$ provides a more balanced form of the $(2,3)$-integer partition.    Thus,  any number  $x$  belonging to the set ${\mathbb{N}_{2,3}}$   can be uniquely represented in one of the forms $2^{n-1}+ 3^{k}x_{1}$ or $2^{n-2} + 3^{k}x_{1}$, where $x_{1}\in {\mathbb{N}_{2,3}}$, $x_{1} <x, k\geq 1$. Applying the same decomposition procedure to $x_{1}$, we obtain the remaining number  $x_{2}$. In general, at the \emph{i}-th stage of the iterative procedure, we get the remaining number $x_{i+1}$, such that $x_{i} = 2^{n_{i}}+3^{k_{i}}x_{i+1}$, where $n_i=\lfloor\log_{2}x_i\rfloor-1$ or $n_i=\lfloor\log_{2}x_i\rfloor-2$. Continue this process recursively until at a certain iteration $t-1$ we obtain $x_{t} = 1$ or  $x_{t} = 2$ (in the last case  $x_{t-1} = 7 = 2^{0} + 3\cdot 2)$.

A lower $(2,3)$-code is defined as any code in the binary alphabet $\{0,1\}$ that can be used to restore the sequence of values  $x_{t}, x_{t-1},\ldots, x_{1}, x$. One of such codes we obtain using the so-called $(\Delta,k)$- approach.

Note that for the unambiguous reconstruction of the number $x$ it is sufficient to keep the sequence of pairs given by the values $\Delta_{i} = \lfloor log _{2}3^{k_{i}}x_{i+1}\rfloor -n_{i}$   and $k_{i}, i = 0,\ldots,t-1$. These pairs  we  obtain at each iteration during decomposition of $x$. For the lower $(2,3)$-representation the following remarkable property holds. The defined above parameter $\Delta_{i}$ can take only three values: $0, 1$ and $2$ \cite{AZ}.

So, with a number $x$ the numerical sequence of pairs is uniquely associated $(\Delta_{0},k_{0}), (\Delta_{1},k_{1}),\ldots,(\Delta_{t-1},k_{t-1})$, where $0 \leq\Delta_{i} \leq 2,  0 < k_{i}$.

For the lower $(2,3)$-encoding, we use the specific binary encoding of pairs. The value $\Delta$  is encoded as follows: $\Delta = 2$ by the symbol 0, $\Delta=1$  by the word  $11$ and $\Delta=0$ by the word $10$. The value $k$ is encoded by the word $1^{k-1}0$ with some exceptions arising due to the selection of a delimiter. In these exceptional cases, the codeword for $k$ is $1^{k}0$.

The codeword of a number $x$ is the sequential concatenation of the corresponding $(\Delta,k)$-groups. For the lower $(2,3)$-code encoding groups are written in the reverse order regarding the way of obtaining them during encoding, $(\Delta_{t-1},k_{t-1}),\ldots,( \Delta _{0},k_{0})$. This allows to perform the decoding from left to right and makes it easier.

Since every $(\Delta,k)$-group, and each codeword ends with the symbol $0$, then the word $0110$ can serve as a delimiter.

To form the delimiter, it is necessary to append  the string  $110$ to the end of some words. If  in a codeword the last group corresponding to the pair $(\Delta_{0}, k_{0})$ takes the form $0110$ or $10110$, i.e.  $k_{0} = 3$ and $\Delta_{0}\neq 1$, then it already contains the delimiter, so there is no need to postfix the string $110$ to the end of a word.

Thus, the $(\Delta,k)$-groups $110$, $0110$, $10110$ are separating ones; if any of them occurs, a codeword ends with it. In a codeword the last group $110$, which is externally appended, does not correspond to any pair $(\Delta,k)$ that take part in the lower $(2,3)$-representation, and has to be ignored during decoding, but groups $0110$ and $10110$ have to be taken into consideration. So, none $(\Delta,k)$-group that corresponds to a pair should not take the form 110, and none $(\Delta,k)$-group except the last one, should not take the forms $0110$ or $10110$. However, codewords of pairs $(\Delta_{i},k_{i})$ received in the lower $(2,3)$-factorization can violate these conditions. Namely, this undesirable situation occurs when:
\begin{enumerate}
\item $\Delta = 1$ and $k = 1$ (then the group 110 is formed);
\item  $\Delta\neq 1$, $k = 3$  and the corresponding $(\Delta,k)$-group is not the last one
    (it is one of the groups $0110$ or $10110$).
    \end{enumerate}

It is easy to check (and this is shown in \cite{AZ}) that for the group $(\Delta_{t-1}, k_{t-1})$, which is written first in a codeword, case 1) is impossible. Therefore, to avoid the undesirable situation mentioned above, instead of $1^{k-1}0$ we encode the value $k$ in a $(\Delta,k)$-group by the string $1^{k}0 $ in such cases:

 $\Delta = 1$ and a $(\Delta,k)$-group is not the first;

 $\Delta\neq 1, k\geq 3$ and a $(\Delta,k)$-group is not the last.

In this way, the constructed prefix code corresponds to the set of positive integers that are coprime with 2 and 3. The number 1, for which the lower $(2,3)$-factorization is empty, corresponds to the shortest codeword  $110$. Together with the last zero of a preceding codeword this sequence forms a delimiter.

By $C^{low}_{2,3}$ we  denote the lower $(2,3)$-code described above.

To encode  an arbitrary positive integer $n$, it is necessary to find the $n$-th number in the ascending  series of numbers that are coprime with 2 and 3. This number equals to $x = 3n-(n\mod 2)-1$. Thus, to encode $n,$ one have to find the lower $(2,3)$-representation of $x$ and encode it.

Table \ref{tab1} shows 15 smallest numbers, their lower $(2,3)$-representations, and the corresponding codewords of the lower $(2,3)$-code.

\begin{table}[!t]
\caption{Lower $(2,3)$-representations and codewords of the first fifteen numbers}
\label{tab1}
\centering
\begin{tabular}{|l|l|l|l|l|l|l|l|}

\hline
$n$ & $x$ & $(\Delta_{0},k_{0})$ & $x_{1}$ & $(\Delta_{1},k_{1})$ & $x_{2}$ & code \\
\hline

\hline
1 & 1 &  &  &  & & 110  \\
\hline

\hline
2 & 5 & 0,1 &1 &  &  & 100 110  \\
\hline

\hline
3 & 7 & 2,1 & 2 &  &  & 00 110  \\
\hline

\hline
4 & 11 & 2,2 & 1 &  &  & 010 110  \\
\hline

\hline
5 & 13&1,2  & 1 & &  &1110 110 \\
\hline

\hline
6 & 17 & 0,2 & 1 &  & & 1010 110 \\
\hline

\hline
7 & 19 & 1,1 & 5 & 0,1 & 1 & 100 1110 110  \\
\hline

\hline
8 & 23 & 0,1 & 5 & 0,1 & 1 & 100 100 110  \\
\hline

\hline
9 &25 & 2,1 & 7 & 2,1 & 2 & 00 00 110  \\
\hline

\hline
10 & 29 & 1,1 & 7 & 2,1 & 2 & 00 1110 110  \\
\hline

\hline
11 & 31 & 2,3 & 1 &  &  & 0110   \\
\hline

\hline
12 & 35 & 1,3 & 1 &  &  & 11110 110  \\
\hline

\hline
13 & 37 & 0,1 & 7 & 2,1 & 2 & 00 100 110 \\
\hline

\hline
14 & 41 & 2,1 & 11 & 2,2 & 1 & 010 00 110  \\
\hline

\hline
15 &43 & 0,3 & 1 &  &  & 10110  \\
\hline

\end{tabular}
\end{table}

As it was mentioned above, the last element in the lower $(2,3)$-representations is the number $x_{t} = 1$ or $x_{t} = 2$. Hence, decoding starts from one of these numbers. Then the  sequence  of numbers $x_{t},\ldots, x_{1}$, $x_{0}=x$ is calculated. It is processed as follows. Using the values $x_{i+1}, \Delta_{i}$ and $k_{i}$ we calculate $n_{i}=\lfloor log_{2}3^{k_{i}}x_{i+1}\rfloor -\Delta_{i}$, and hence we can obtain $x_{i}=2^{n_{i}}+3^{k_{i}}x_{i+1}$. Note that $x_{t}=2$ if and only if $\Delta_{t-1}=2$ and $k_{t-1}=1$; in other cases $x_{t}=1$ \cite{AZ}. Thus, there is no ambiguity at the starting point of the decoding procedure.

\section{Code $D_{2}$}

The existence of  a delimiter for the code $C^{low}_{2,3}$ means that this code is prefix-free. However, it is not complete, i.e. the set of its codewords can be expanded while its UD property will not be lost. To demonstrate that, we construct a prefix code that contains all the codewords from $C^{low}_{2,3}$,  and some more.

This code  is quite simple to define. It consists of the word $110$, and all other binary words that do not start with the string 110, ends with the sequence $0110$ and do not contain this sequence as a substring in other places. We denote this code by $D_{2}$. The number 2 in the code notation  indicates that its delimiter contains 2 consecutive ones.

Obviously, the code $D_{2}$ contains all the codewords of the code $C^{low}_{2,3}$ and has the same delimiter $0110$ as the code $C^{low}_{2,3}$.

 Each portion of concatenated codewords from  $D_{2}$  ends with the delimiter string $0110$ that makes it possible  to unambiguously determine the beginning of a new codeword in the flow of codewords.

 This also provides synchronizability of the code. In case of errors occur a receiver has only to identify the first delimiter string $0110$ to renew the code parsing. But in some cases it cannot unambiguously identify the delimiter suffix $110$ as the single codeword.

The example of a word belonging to the code $D_{2}$, but not to $C^{low}_{2,3}$,  is $100 00 110$. If we apply the $(2,3)$-decoding procedure to this string, we obtain the number 17. However, as Table \ref{tab1} shows, the codeword for $17$ is $1010 110$.

Thus, the code $C^{low}_{2,3}$ is not complete. By the contrast, the code $D_{2}$ is complete, as a representative of a wider class of complete codes that will be defined and investigated in the following sections.

\section {Splittable codes}

In the lower $(2,3)$-integer representation, we use sequences of $(\Delta,k)$-pairs. Let us change the order of $\Delta$  and $k$ inside pairs. In this way, the dual sequence of $(k,\Delta)$-pairs $(k_{i},\Delta_{i})$, where $k_{i}$ is an arbitrary positive integer, and $\Delta_{i}$ takes the same values 0, 1 or 2, can also be associated with a  number.

Apart from the above-mentioned, this representation allows other binary prefix encodings including the following. We represent the value $k$ as the word $0^{k-1}1$ in the unary numeration system with 1 as a separator and the value  $\Delta$ in the form $1^{\Delta}0$. The concatenation of codewords corresponding to $k_{i}$ and $\Delta_{i}$   respectively constitutes a
$(k,\Delta)$-group. The codeword of a $(k,\Delta)$-sequence is formed by the concatenation of corresponding $(k,\Delta)$-groups appended by the delimiter string $1111$. It is obvious that in the concatenation of $(k,\Delta)$-groups obtained through the $(2,3)$-decomposition that word does not occur.

In the lower $(2,3)$-integer representation,  not all possible $(k,\Delta)$-sequences are valid. Let us abstract ourselves from the semantics of values $k$ and $\Delta$, as parameters of the lower $(2,3)$-factorization. Using the defined above atomic encoding of $(k,\Delta)$-pairs we consider encoding all possible sequences of $(k,\Delta) $-pairs   $(k_{1},\Delta_{1})(k_{2},\Delta_{2})\ldots(k_{t},\Delta_{t})$, where the following restrictions hold: $0 \leq \Delta_{i}\leq 2$, $0<k$. It is easy to see that the obtained set of codewords is nothing more than the code Fib4, named in \cite{AF} as the code $C_{1}$ of the order 4.

In this way varying upper bounds for values $\Delta$, $0\leq \Delta_{i}\leq m$, and, respectively, the quantity of ones in a code delimiter we obtain different Fibonacci codes. So, if $\Delta$  can take only one value (which is encoded by "0") and the delimiter consists of two ones, then we obtain the code Fib2. If  $\Delta$ can take two values, which we encode by words "0" and "10", then the delimiter consists of three consecutive ones, and we have the code Fib3. Overall, in Fibonacci codes a restriction on the set of $\Delta$-values naturally predetermines a delimiter.  If $\Delta$  can take no more than $m$ different values, then the delimiter is the run of $m+1$ ones.

Thus, we can assume that the lower $(2,3)$-code, the popular Fibonacci codes and possibly some others can be viewed as the different realizations of a more general method of number encoding based on encoding sequences of ordered integer pairs with limitations on one of their components.

From a practical point of view, it is also important that a code contains a sufficient number of short words. This means that if we consider a code with delimiters, the delimiters or their prefix  parts should be included in some short sequences of $(\Delta,k)$ or $(k,\Delta )$-groups. The longer codewords can contain these shorter words as suffixes and thus we may not consider  delimiters apart from codes of $(\Delta,k)$ (or $(k,\Delta)$)-sequences. Summarizing all the above mentioned, we come to the following definition of $(\Delta,k)$-codes.

\begin{defi} Let $\textbf{S}$ be a given set of sequences of $(\Delta,k)$-pairs, where $\Delta$ is a non-negative integer that does not exceed some constant $d$, and $k$ can be any positive natural number. A $(\Delta,k)$-code of $\textbf{S}$ is the set of binary words that satisfy the following conditions:

\begin{itemize}{\IEEEsetlabelwidth{(iii)}}
\item[(i)] values  $\Delta$ and $k$ are encoded by separate independent prefix encoding functions $\varphi_{1}$ and $\varphi_{2}$ respectfully;

\item[(ii)]	the encoding of a  $(\Delta,k)$-pair is defined as the concatenation $\varphi_{1}(\Delta)\varphi_{2}(k)$, which we call a $(\Delta,k)$-group;

\item[(iii)] the codeword of a $(\Delta,k)$-sequence from $\textbf{S}$ is the sequential
 concatenation of the corresponding $(\Delta,k)$-groups.
\end{itemize}
\end{defi}

A $(\Delta ,k)$-code is any set of binary words that can be interpreted as a $(\Delta ,k)$-code for some set $\textbf{S}$ of $(\Delta ,k)$-sequences.

Thus, to set  a $(\Delta ,k)$-code it is necessary to specify a set $\textbf{S}$ of $(\Delta ,k)$-sequences and to choose well defined basic encodings of  $(\Delta ,k)$-pairs.

In what  follows, we consider only codes, where a set $\textbf{S}$ is the set of all possible $(\Delta ,k)$-sequences. In general, like in the case of  $(2,3)$-codes, a basic set $\textbf{S}$ could be a subset of all $(\Delta ,k)$-sequences.

The definition of a $(k, \Delta)$-code is similar to that given above by changing $(\Delta ,k)$ by $(k, \Delta )$-pairs.

We call both the $(\Delta ,k)$ and $(k,\Delta)$-codes  splittable codes.

The important property of splittable codes is that any codeword, including a delimiter, consists of a whole number of $(\Delta ,k)$ (respectively  $(k, \Delta)$)-groups. This structural regularity can also be used as an element of proving technique in establishing important code properties, such as completeness, universality, and density.

 As shown above, the codewords of Fibonacci codes can be represented as sequences of $(k,\Delta)$-groups, which are externally supplemented by a delimiter. Interestingly, that using specific encodings  of  $k$ and $\Delta$, these codewords can be interpreted as the sequences consisting of a whole number of $(k,\Delta)$-groups even with a delimiter. Nevertheless, they cannot be given as the sequences  of $( \Delta,k)$-groups.

\begin{theorem}
Any Fibonacci code Fib$m$ is a $(k,\Delta)$-code, but not a $(\Delta ,k)$-code.
\end{theorem}

\begin{IEEEproof} Consider a $(k,\Delta)$-pair, where $k$ could be any positive integer,  and $\Delta$  can have only  $m$ different values,  $0 \leq\Delta < m$. Let us encode $k$ by the string $0^{k-1}1$, which comprises $k-1$ zeros. Values of $\Delta$   we encode by $m$ strings: $0, 10,\ldots, 1^{m-2} 0$, which contain runs up to $m-2$ ones, and the string $1^{m-1}$  corresponding to the value  $m-1$.

Using this encoding we prove the first part of the theorem statement by induction on the codeword length.

Let  $\alpha$ be a codeword from Fib$m$. The minimal possible length of $\alpha$ is equal to $m$. If that is so, $\alpha = 1^{m}=11^{m-1}$.    This string corresponds to the $(k,\Delta)$-pair $(1, m-1)$.

Suppose that the statement of the theorem holds for all codewords having lengths less or equal to some integer $t$, $t \geq m$.  Assume that the length of $\alpha$  is $t+1$.

If $\alpha$  starts with 1, then $\alpha$  can be represented in the form $\alpha=1^{i}0\beta=11^{i-1}0\beta, 0<i<m$.  The prefix $11^{i-1}0$  corresponds to the  $(k,\Delta)$-pair $(1, i-1)$. The shorter string $\beta$  also belongs to Fib$m$. Thus, by the inductive assumption $\beta$   comprises an integral number  of $(k,\Delta)$-groups.

Consider the case when $\alpha$  starts with $0, \alpha=0^{i}1\beta, i>0$.   If $\beta$  is the suffix of the form $1^{m-1}$  then $\alpha=0^{i}11^{m-1}$, and that corresponds to the $(k,\Delta)$-pair $(i+1, m-1)$.

In another case,  $\beta$ is a string of the form $\beta=1^{j}0\gamma, 0\leq j < m-1, \gamma \in$ Fib\emph{m}. This gives the representation form $\alpha=0^{i}11^{j}0\gamma$. The prefix part $0^{i}11^{j}0 $   is the codeword corresponding to the $(k,\Delta)$-pair $(i+1, j.)$ By the inductive assumption the string $\gamma$  contains a  whole number of $(k,\Delta)$-groups. Hence, $\alpha$  corresponds to some  $(k,\Delta)$-sequence. By induction the first part of Theorem 1 is proved.

Consider the second part of the theorem. Suppose, to the contrary, that Fib$m$  is a  $(\Delta,k)$-code with some prefix encoding functions $\varphi_{1}$ for  $\Delta$-values and $\varphi_{2}$   for $k$-values.

For any integer $k$  the codeword $0^{k}1^{m}$  belongs to Fib$m$. On the other hand, the lengths of codewords corresponding to $\Delta$ values are restricted. It follows that there exists the value $\Delta^{\prime}$ such that $\varphi_{1}(\Delta^{\prime})=0^{s}$  for some integer $s>0$.

Consider the word $0^{s}1^{m}$. The prefix property of the encoding $\varphi_{1}$ implies that there are no other codes of $\Delta$ of the form $0^{r}, r<s$. It follows that there exists some value  $k^{\prime}$ such that $\varphi_{2}(k^{\prime})=1^{t}, t>0$,  and $\varphi_{1}(\Delta^{\prime})\varphi_{2}(k^{\prime})$  is the first $(\Delta, k)$-group for the string $0^{s}1^{m}$.

Consider the string $1^{m}$.  It also belongs to Fib$m$. By our assumption, some  $(\Delta,k)$-groups constitute the representation
$1^{m}=\varphi_{1}(\Delta_{1})\varphi_{2}(k_{1})...\varphi_{1}(\Delta_{n})\varphi_{2}(k_{n})$.

The prefix property of encodings $\varphi_{1}$ and $\varphi_{2}$  implies that
$\Delta_{1}=\Delta_{2}=\ldots=\Delta_{n},\ k^{\prime}=k_{1}=k_{2}=\ldots=k_{n}, $
$\varphi_{1}(\Delta_{1}) = 1^{r}, r>0, \varphi_{2}(k_{1})=1^{t}, t>0$.

It immediately follows that the inequality $t<m$ holds.

Thus, from the consideration of the string $0^{s}1^{m}$  we conclude that the non-empty string $1^{m-t}$  consists of a whole number of identical $(\Delta, k)$-groups. Each of them corresponds to the pair $(\Delta_{1}, k_{1})$.

The string $1^{m}$  can be represented in the form $1^{m}=1^{m-t}1^{t}$. It follows that the string $1^{t}$  should be represented using an integral quantity of identical $(\Delta, k)$-groups corresponding to the encoding $\varphi_{1}(\Delta_{1})\varphi_{2}(k_{1})=1^{r+t}, r>0$. This contradiction concludes the proof.
\end{IEEEproof}

For Fibonacci codes considered as $(k, \Delta )$-codes we use the unary encoding of parameters $k$ and $\Delta$. Note that when we use splittable codes for data compression, then they can be more effective, if the average codeword length is shorter. From this perspective, the encoding of parameters $k$ and $\Delta$  in the unary numeration system is not economical. More economical, for example, is the truncated binary encoding of the values $\Delta$ and $k$. However, for the parameter $k$ such encoding is impossible since the set of its values is unlimited. Nevertheless, the truncated binary encoding can be applied to encode the values of the parameter $\Delta$.

Concerning the parameter $k$, there are only two unary prefix encodings $0^{k-1}1$  or $1^{k-1}0.$  Theoretically, other prefix encodings, such as Elias codes \cite{El} can be used for encoding $k.$ However, in applications of splittable codes to text compression, the probability distribution of  $k$-values is geometric,  and unary codes are the most effective for this kind of distribution.

The Golomb codes \cite{Golomb} completely correspond to the principles described above. Those are ones of the simplest $(k,\Delta )$-codes, where each codeword consists of one $(k, \Delta )$-group.

If we consider more complex codes, which codewords can contain several $(\Delta,k)$ or $(k,\Delta)$-groups, then certain groups should be considered as terminating in a codeword, i.e. separating ones. We note that due to the unary encoding of the parameter $k$, the last bit of any $(\Delta,k)$-group always has the same value, say zero. Therefore, to endow a splittable code with the feature of instantaneous separation, it is suitable to construct a code from $(\Delta,k)$-, but not $(k,\Delta)$-groups,  predetermining a delimiter as $0\alpha 0$, where  $\alpha0$-is a separating group, and zero in front of it is the last symbol of the previous group. If  we encode $\Delta$  in the binary form, then $(k,\Delta)$-groups will not have such properties, because they can begin and end with zero as well as with one. This complicates finding the place that matches a delimiter.

However, the more important advantage of $(\Delta,k)$-codes over $(k, \Delta)$-codes is the possibility to form short codewords that do not contain a whole delimiter. For example, they can consist of a separating group of the form   $\alpha0$, while the delimiter takes the form  $0\alpha0$. Longer delimiters provide the better asymptotic density of a code, while  short codewords enable us to organize efficient compression for relatively small alphabet  sizes. Thus, for example, the considered above code $D_{2},$  it will be proved further that it is a $(\Delta,k)$-code, contains  the word $110$, although the sequence $0110$ is the code delimiter. As will be shown, it has a higher asymptotic density than the code Fib3, and only slightly inferior in the efficiency of compressing texts with small alphabets.

\section{Multi-delimiter codes}

One of the families of efficient $(\Delta ,k)$-codes can be obtained by using several delimiters of the form $01^m0$ in one code.  The remaining part of this presentation deals completely with the investigation of these codes.

Let $\mathcal{M}=\{m_{1},\ldots,m_{t}\}$  be a set of integers, given in the ascending order, $0 < m_{1}<\ldots< m_{t}$.

\begin{defi} The multi-delimiter code $D_{m_1,\ldots,m_t}$ consists of  all the words of the form  $1^{m_{i}}0, i=1,\ldots,t$ and all other words that meet the following requirements:

\begin{itemize}{\IEEEsetlabelwidth{(iii)}}
\item[(i)] for any $m_{i}\in\mathcal{M}$ a word does not start with a sequence $1^{m_{i}}0$;

\item[(ii)] a word  ends with the suffix $01^{m_{i}}0$  for some  $m_{i}\in\mathcal{M}$;

\item[(iii)] for any $m_{i}\in\mathcal{M}$ a word cannot contain the sequence $01^{m_{i}}0$  anywhere, except a suffix.
\end{itemize}
\end{defi}

The given definition implies that code delimiters in  $D_{m_1,\ldots,m_t}$ are sequences of the form $01^{m_{i}}0$.  However, the code also contains shorter words of the form $1^{m_{i}}0$, which form the delimiter together with the ending zero of a preceding codeword.

Evidently, any multi-delimiter code is prefix-free and thus UD.

Table \ref{tab2} shows examples of multi-delimiter codewords. This table lists all codewords of lengths not longer than 7 of different multi-delimiter codes and, for comparison, Fibonacci codes Fib2 and Fib3.

The codes $D_{2,3}$ and $D_{2,3,4}$ with 2 and 3 delimiters respectfully contain many more short codewords than both the Fibonacci code Fib3 and the one-delimiter code $D_{2}$. However, as it will be demonstrated in the following, the asymptotic density of these codes is lower.

Overall, codes with more delimiters have worse asymptotic density,  but contain a larger quantity of short codewords. This regularity is related also to the lengths of delimiters: the shorter they are, the larger quantity of short words a code contains.

For natural language text compression, the most effective seems to be codes with the shortest  delimiter having two ones, which we will thoroughly examine.

\begin{table*}[!t]
\caption{Sample codeword sets of some multi-delimiter and Fibonacci codes}
\label{tab2}
\centering
\begin{tabular}{l@{\extracolsep{2mm}}lllllll}
\hline\noalign{\smallskip}
Index&Fib2&$D_1$&$D_{1,2}$&Fib3&$D_2$&$D_{2,3}$&$D_{2,3,4}$\\
\hline\noalign{\smallskip}
1&11&10&10&111&110&110&110\\
\cline{2-8}
2&011&010&010&0111&0110&0110&0110\\
\cline{2-3}\cline{5-6}
3&0011&0010&110&00111&00110&1110&1110\\
\cline{3-4}\cline{7-8}
4&1011&00010&0010&10111&10110&00110&00110\\
\cline{2-2}\cline{5-6}
5&00011&11010&0110&000111&000110&10110&10110\\
\cline{3-4}
6&01011&000010&00010&010111&010110&01110&01110\\
\cline{7-7}
7&10011&011010&00110&100111&100110&000110&11110\\
\cline{2-2}\cline{4-4}\cline{6-6}\cline{8-8}
8&000011&110010&000010&110111&0000110&010110&000110\\
\cline{5-5}
9&001011&111010&000110&0000111&0010110&100110&010110\\
\cline{3-3}
10&010011&0000010&111010&0010111&0100110&001110&100110\\
\cline{4-4}
11&100011&0011010&0000010&0100111&1000110&101110&001110\\
\cline{7-7}
12&101011&0110010&0000110&1000111&1010110&0000110&101110\\
\cline{2-2}
13&0000011&1100010&0111010&1010111&1110110&0010110&011110\\
\cline{6-6}\cline{8-8}
14&0001011&0111010&1110010&0110111&&0100110&0000110\\
15&0010011&1110010&1110110&1100111&&1000110&0010110\\
\cline{5-5}
16&0100011&1111010&1111010&&&1010110&0100110\\
\cline{3-4}
17&1000011&&&&&0001110&1000110\\
18&0101011&&&&&0101110&1010110\\
19&1001011&&&&&1001110&0001110\\
\cline{7-7}
20&1010011&&&&&&0101110\\
\cline{2-2}
21&&&&&&&1001110\\
22&&&&&&&0011110\\
23&&&&&&&1011110\\
\cline{8-8}
\end{tabular}
\end{table*}

Now we demonstrate that multi-delimiter codes belong to the class of splittable codes.

\begin{theorem}
Any multi-delimiter code $D_{m_1,\ldots,m_t}$ is a $(\Delta,k)$-code.
\end{theorem}

\begin{IEEEproof} We need to set some positive integer that cannot be exceeded by the value of $\Delta$ and construct prefix encodings for $\Delta$ and $k$ so that any codeword of  $D_{m_1,\ldots,m_t}$ comprises a whole number of $(\Delta, k)$-groups.

Let $d$ be  some fixed non-negative integer satisfying inequalities $0 \leq d<m_{1}$. The parameter $\Delta$ ranges from $0$ to  $2^{d}+1$. We encode these values by the symbol $0$ and all binary words  of the length $d+1$ with the fixed first symbol $1$. The value of the parameter $k,$ which can be any positive integer, is encoded by the word $1^{k-1}0$. Evidently, these encodings of values $\Delta$ and $k$ are prefix-frree.

Consider a word $1^{r}0$,  where $r\geq m_{1}$. This word can be represented in the form $1^{r}0 = 1^{d+1}1^{r-d-1}0$. The inequality $r\geq m_{1}$ and the choice of $d$  implies that $r\geq d+1$. It follows that $1^{r}0$  corresponds to the $(\Delta, k)$-pair with $\Delta$  encoded by $1^{d+1}$ and $k=r-d>0$ and any word $\alpha\in D_{m_1,\ldots,m_t}$ of the form $1^r0$ represents some $(\Delta, k)$-group.

 Note that for any binary word  $\alpha$ of the length exceeding $d$ and containing zeros in its representation it is possible to choose a prefix, such that it can be interpreted both as a codeword of some value $\Delta$, and as a codeword of some value $k$.  Indeed, if $\alpha$  starts with $0$ then this symbol can be interpreted as corresponding to  $\Delta =0$ or $k=1$.  If $\alpha$  starts with 1 then $\alpha = 1^{r}0\beta$,  where $r>0 $ and $\beta$  is the binary word. The prefix $1^{r}0$  can be interpreted as the codeword of the value $k=r+1$. But, also it is possible to choose the prefix of $\alpha$  having the length $d+1$, which corresponds to some value of  $\Delta$.

Now, suppose that $\alpha\in D_{m_1,\ldots,m_t}$ and it does not have the form $1^r0$.  Let us consider parsing the codeword $\alpha$  from left to right sequentially extracting corresponding $(\Delta,k)$-groups until it is possible. As the result, we make partitioning of $\alpha$  on a whole number of $(\Delta,k)$-groups or we obtain a remainder that is not capable of containing a whole number of $(\Delta,k)$-groups.

In the first case we obtain the desirable partitioning of $\alpha$ on an integral number of $(\Delta,k)$-groups.

Consider the case of obtaining a remainder. Let us examine how under this procedure the ending of a codeword is processed. The suffix of  a codeword  has the form $01^{m_{i}}0$  and contains at least $m_{1}$ ones. The first bit $"0"$ of that suffix either can be  the ending of some codeword of $k$ or can belong to a codeword of $\Delta$. In the first case, at the last iteration we obtain the  residue $1^{m_{i}}0$ with no less than $m_{1}$ ones that, as shown above, is a $(\Delta,k)$-group. In the second case, we note that the codeword of $\Delta$  comprises no more than $m_{1}$ bits and after its extraction we obtain the remaining sequence of the form $1\ldots10$, which represents a particular value of $k$. Thus, the situation when at the last iteration we obtain a remainder, which is not capable of containing a whole $(\Delta,k)$-group, is impossible.
\end{IEEEproof}

Note that Theorem 2 holds for any values $d$ that satisfy the inequalities $0\leq d<m_{1}$. In the sequel to further simplify considerations, we presume that $d=0$, i.e. the code of a $\Delta$-value  comprises one bit.

Note that although in the code $D_{2}$  we used the encoding of three possible values of  $\Delta$,  which corresponds to the value $d=1$, all words of that code can be also represented as  $(\Delta,k)$-groups with a single-bit encoding of $\Delta.$

\begin{theorem}
Any code $D_{m_1,\ldots,m_t}$ is complete.
\end{theorem}

\begin{IEEEproof} A necessary and sufficient condition for a code $C$ to be complete is given by the Kraft-Macmillan equality: $\sum\limits_{c\in C}2^{-|c|}=1$.
By  $f_{n}$ denote the number of codewords of the length $n$.  This equality can be rewritten as:

\begin{equation}
\label{q1}
\sum_{n=1}^\infty2^{-n}f_n=1
\end{equation}

Consider the  multi-delimiter code $D_{m_1,\ldots,m_t}$.

Theorem 2 allows us to choose the one-bit encoding for $\Delta$, and $k$ is encoded by $1^{k-1}0$.

For any $n \geq 2$ there exist two $(\Delta,k)$-groups of length $n$: $1^{n-1}0$ and $01^{n-2}0$. Among all of them $(\Delta,k)$-groups that include $m_{i}$ ones, $i=1,\ldots, t$, are terminal, i.e. they can occur only at the end of a codeword. Thus, for the code $D_{m_1,\ldots,m_t}$ there are  $2t$ terminal groups having  lengths  $m_{1}+1,m_{1}+2,\ldots, m_{t}+1,m_{t}+2$.

By $T_{n}$  denote the number of terminal groups of the length $n$. Evidently, $T_{n}$ equals to the number of occurrences of $n$ in the set $\{m_{1}+1,m_{1}+2,\ldots, m_{t}+1,m_{t}+2\}$.
 This number can be equal to $0, 1$ or $2$. The number of non-terminal groups of length $n$ equals to $2-T_{n}$.

Consider the codewords of the length $n$ that contain at least two $(\Delta ,k)$-groups. Each such word can be obtained by prepending its first non-terminal $(\Delta ,k)$-group to a shorter codeword. On the other hand, prepending an arbitrary non-terminal group to any codeword forms a longer codeword. If the codeword contains only one $(\Delta ,k)$-group, then this group is terminal. Thus, taking into account that the length of the shortest $(\Delta ,k)$-group is 2, we obtain the following recurrent formula for calculating the number of codewords of the length $n$:

\begin{eqnarray}
\label{q2}
f_n=T_{n}+\sum_{k=0}^{n-2}(2-T_{n-k})f_{k} = \nonumber\\
=T_{n}+2(f_{n-2}+f_{n-3}+ \cdots)- \nonumber\\
-f_{n-(m_1+1)}-\cdots-f_{n-(m_t+1)}-\nonumber\\
-f_{n-(m_1+2)}-\cdots-f_{n-(m_t+2)}
\end{eqnarray}

Let us apply this formula to calculate $f_{n-1}$:

\begin{eqnarray}
\label{q3}
f_{n-1}=T_{n-1}+\sum_{k=0}^{n-3}(2-T_{n-1-k})f_{k}=\nonumber \\
=T_{n-1}+2(f_{n-3}+f_{n-4}+\cdots)-\nonumber \\
-f_{n-(m_1+2)}-\cdots-f_{n-(m_t+2)}- \nonumber \\
-f_{n-(m_1+3)}-\ldots-f_{n-(m_t+3)}
\end{eqnarray}

Find the right part of (3) in (2) and change it to $f_{n-1}$:

\begin{eqnarray}
\label{q3}
f_{n}=T_{n}-T_{n-1}+2f_{n-2}+f_{n-1}-\nonumber \\
f_{n-m_{1}-1}-\cdots -f_{n-m_{t}-1}+\nonumber \\
+f_{n-m_{1}-3}+\cdots+f_{n-m_{t}-3}
\end{eqnarray}

Denoting the  left part of (1) by $s$ and taking into account that
 $f_{0}=f_{-1}=\cdots=0$, for any $p>0$ we have the following equalities:
$\sum_{n=1}^\infty2^{-n}f_{n-p}=2^{-p}\sum_{n=1}^\infty2^{-(n-p)}f_{n-p}=s2^{-p}$.

Taking them into consideration and substituting expression (4) in (1), we obtain the following:

\begin{eqnarray}
\label{q5}
s=\sum_{n=1}^\infty2^{-n}f_n=\sum_{n=1}^\infty2^{-n}(T_{n}-T_{n-1}+f_{n-1}+\nonumber\\
2f_{n-2}-f_{n-(m_{1}+1)}-\cdots-f_{n-(m_{t}+1)}+\nonumber\\
+ f_{n-(m_{1}+3)}+\cdots+f_{n-(m_{t}+3)}=\nonumber\\
=\sum_{n=1}^\infty2^{-n}T_{n}-\frac{1}{2}\sum_{n=1}^\infty2^{-(n-1)}T_{n-1}+\nonumber\\
+s(\frac{1}{2}+\frac{1}{2}-2^{-m_{1}-1}-\cdots- 2^{-m_{t}-1}+\nonumber\\
+2^{-m_{1}-3}+\cdots+2^{-m_{t}-3})
\end{eqnarray}

Taking into account that $2^{-m_i-3}-2^{-m_i-1}=-3\cdot2^{-m_i-3}$ for any $i$, $\sum_{n=1}^\infty2^{-n}T_n=\sum_{n=1}^\infty2^{-(n-1)}T_{n-1}$ and cancelling out $s$ in both parts of (\ref{q5}) we obtain the following formula.

\begin{equation}
\label{q6}
3s\sum_{i=1}^t2^{-m_i-3}=\frac{1}{2}\sum_{n=1}^\infty2^{-n}T_{n}
\end{equation}

Since the lengths of terminal $(\Delta ,k)$-groups are $m_{1}+1 , m_{1}+2 ,\ldots, m_{t}+1, m_{t}+2$, the equality
$$
\sum_{n=1}^\infty2^{-n}T_{n}= \sum_{i=1}^t2^{-m_i-1}+2^{-m_i-2}=\\
\frac{3}{4}\sum_{i=1}^t2^{-m_i}
$$
is satisfied.

Therefore, equality (6) takes the form

$$\frac{3}{8}s\sum_{i=1}^t2^{-m_{i}}= \frac{3}{8}\sum_{i=1}^t 2^{-m_{i}}$$

That implies the condition $s = 1$.
\end{IEEEproof}

Also the  $(\Delta ,k)$-structure of multi-delimiter codes enables us to prove another important feature, universality, but we give the simpler proof based on encoding integers.

\section{Encoding integers}

We define a multi-delimiter code as a set of words. There exists a simple bijection between the set of natural numbers and the set of codewords of any multi-delimiter code. Thus, it enables us to encode integers by codewords of these codes.

Let $\mathcal{M}=\{m_{1},\ldots, m_{t}\}$  be the set of parameters of the code $D_{m_1,\ldots,m_t}$. By   ${\mathbb{N}_{\mathcal{M}}}=\{j_{1},j_{2},...\}$ denote the ascending sequence of all natural numbers that does not belong to $\mathcal{M}$.

\emph{Example}.  Let $\mathcal{M}=\{2,5\}$. This gives the set ${\mathbb{N}_{\mathcal{M}}}=\{1,3,4,6,7,8,...\}.$

By $\varphi_{\mathcal{M}}(i)$  denote the function $\varphi_{\mathcal{M}}(i)=j_{i}$, $j_{i}\in {\mathbb{N}_{\mathcal{M}}}$  as defined above.

It is easy to see that the function $\varphi_{\mathcal{M}}$  is a bijective mapping of the set of natural numbers onto ${\mathbb{N}_{\mathcal{M}}}$.   Evidently, this function and the inverse function $\varphi_{\mathcal{M}}^{-1}$ can be constructively implemented by simple one cycle iterative procedures.

The main idea of encoding integers by the code $D_{m_1,\ldots,m_t}$ is as follows. We scan the binary representation of an integer from left to right. During this scan each internal isolated group of  $i$  consecutive $1s$ is changed to $\varphi_{\mathcal{M}}(i)$  $1s$. This way we exclude the appearance of delimiters inside a codeword. In decoding we change internal isolated groups of $j$  consecutive $1s$ to the similar groups of $\varphi_{\mathcal{M}}^{-1}(j)$  ones. Detailed description of the encoding procedure is as follows.

\emph{Bitwise Integer Encoding Algorithm}.

\emph{Input}: $x=x_{n}x_{n-1}...x_{0}$, $x_{i}\in \{0,1\}, x_{n}=1;$

\emph{Result}: a codeword from $D_{m_1,\ldots,m_t}$.

\begin{enumerate}
\item $x\leftarrow x-2^n$, i.e. extract the most significant  bit of the number $x$, which is always 1.

\item If  $x=0$, append the sequence $1^{m_{1}}0$  to the string $x_{n-1}...x_{0}$, which contains only zeros or empty. \emph{Result} $\leftarrow x_{n-1}...x_{0}1^{m_{1}}0$.  Stop.

\item If the binary representation of $x$ takes the form of a string  $0^{r}1^{m_{i}}0, r\geq0, m_{i}\in \mathcal{M}, i>1,$ then \emph{Result} $\leftarrow x$.  Stop.

\item In the string $x$ replace each isolated group of $i$ consecutive $1s$ with the group of $\varphi_{\mathcal{M}}(i)$ consecutive $1s$ except its occurrence as a suffix of the form  $01^{m_{i}}0, i>1. $ Assign this new value to $x$.

\item If the word ends with a  sequence $01^{m_{i}}0, i>1$, then \emph{Result} $\leftarrow x.$  Stop.

\item Append the string $01^{m_{1}}0$ to the right end of the word. Assign this new value to $x$.
 \emph{Result} $\leftarrow x$.   Stop.
\end{enumerate}

According to this algorithm, if $x\neq 2^{n}$, the delimiter $01^{m_{1}}0$  with $m_{1}$ ones is attributed to a codeword externally, and therefore it should be deleted during the process of decoding, while the delimiters of a form   $01^{m_{i}}0, i>1$ are informative parts of codewords and  they must be processed during the decoding. If $x = 2^{n}$, the last $m_{1}+1$ bits of the form $1^{m_{1}}0$ must be deleted.

\emph{Bitwise Decoding Algorithm}.

\emph{Input}: a codeword $y\in D_{m_1,\ldots,m_t}$.

\emph{Result}: an integer given in the binary form.

\begin{enumerate}
\item If the codeword $y$ is of the form $0^{p}1^{m_{1}}0$, where $p\geq0$,  extract the last $m_{1}+1$ bits and go to step 4.

\item If the codeword $y$ ends with the sequence $01^{m_{1}}0$, extract the last $m_{1}+2$ bits. Assign this new value to $y$.

\item In the string $y$ replace each isolated group of $i$ consecutive $1s$, where $i\in\mathcal{M}$, with the group of $\varphi_{\mathcal{M}}^{-1}(i)$  consecutive $1s$. Assign this new value to $y$.

\item Prepend the symbol 1 to the beginning of $y$. \emph{Result} $\leftarrow y.$  Stop.
\end{enumerate}

The following lemma gives an upper bound for the length of a  multi-delimiter codeword.

\begin{lemma}
Let $D_{m_1,\ldots,m_t}$ be a multi-delimiter code, $c_{i}$  be the codeword of an integer $i$ obtained by the encoding algorithm given above. The length of $c_{i}$ satisfies the following upper bound: $|c_{i}|\leq \frac{1}{2} t \log _{2}i + m_{1}+2$.
\end{lemma}

\begin{IEEEproof} The encoding procedure that transforms a number $i$ given in binary form into the  corresponding codeword of the code $D_{m_1,\ldots,m_t}$ can enlarge each internal isolated group of consecutive $1s$  maximum  on $t$  ones. The quantity of such groups does not exceed  $\frac{1}{2}\log_{2}i$. To some binary words the delimiter $01^{m_{1}}0$  could be externally appended. Therefore, the length of the codeword for $i$ is upper bounded by the value $\frac{1}{2} t\log_{2}i + m_{1} + 2$.
\end{IEEEproof}

Now we are ready to prove that any multi-delimiter code is universal.

The concept of universality was introduced by P. Elias \cite{El}. This notion reflects the property of  prefix sets to be nearly optimal codes for data sources with any given probability distribution function.

A set of the codewords of lengths $l_{i} (l_{1}\leq l_{2}\leq\ldots)$ is called universal, if there exists a constant $K$, such that for any finite distribution of probabilities  $P = (p_{1},\ldots,p_{n})$, where $p_{1}\geq p_{2}\geq\ldots$, the following inequality holds

\begin{equation}
\label{q6}
\sum_{i=1}^n l_{i}p_{i}\leq K \cdot \max (1, E (P)),
\end{equation}

where $E(P)=-\sum_{i=1}^n p_{i}\log_{2} p_{i}$ is the  entropy of distribution $P$, and $K$ is a constant independent of $P$.

\begin{theorem}
Any multi-delimiter code  $D_{m_1,\ldots,m_t}$ is universal.
\end{theorem}

\begin{IEEEproof} Like in Lemma 1, by $c_{i}$ denote the codeword in $D_{m_1,\ldots,m_t}$ corresponding  to the integer $i$. Let us sort codewords from $D_{m_1,\ldots,m_t}$ in the ascending order of their bit lengths, $a_{1}, a_{2},\ldots$. Map them to symbols of the input alphabet sorted in the descending order of their probabilities.

We claim that the length of any word $a_{i}$ also satisfies the length upper bound for $|c_{i}|$  given by Lemma 1.

Indeed, consider the set $\{c_{1},c_{2},\ldots ,c_{i}\}.$ Obviously, each of its elements  satisfies that upper bound. In the sequence $a_{1},{a_{2}\ldots}$ at least one element, say $c_{j}, 1\leq j \leq i,$ occupies the place $k$ such that $k\geq i, a_{k}= c_{j}.$ This implies $|a_{i}|\leq |a_{k}| = |c_{j}|.$ It follows that $|a_{i}|$ satisfies the upper bound for $|c_{i}|,$ which is equal to $\frac{1}{2} t \ log _{2}i + m_{1}+2$ as Lemma 1 stated.

The sequence $a_{1},a_{2},\ldots$ can be considered as a new encoding of natural numbers.
To conclude the proof it remains only to apply the general Lemma 6 by Apostolico and Fraenkel taken from \cite{AF}: "Let  $\psi$  be a binary representation such that $|\psi(k)|\leq c_{1}+c_{2}\log k$ $(k\in \mathbb{Z}^{+})$, where $c_{1}$ and $c_{2}$  are constants and $c_{2} > 0$.  Let $p_{k}$ be the probability to meet $k$. If $p_{1}\geq p_{2}\geq \ldots\geq p_{n}$, $\sum p_{i}\leq 1$ then $\psi$  is universal".
\end{IEEEproof}

\section{Byte aligned algorithms}
The considered above encoding and decoding algorithms are bitwise, and therefore they are quite slow. We can construct accelerated algorithms that process bytes.  Since decoding is performed in real time more often than encoding and in general lasts longer, acceleration of decoding is a more important task we focus on.

The general idea of the byte aligned decoding algorithm is similar to that one described in \cite{Kl1} for the Fibonacci codes. At the \emph{i}-th iteration of this algorithm, a whole number of bytes of encoded text is read out. We denote this portion of text by $u_{i}$. Assume that $u_{i}$ has the form $s_{i}E(w_{1}^i),\ldots,E(w_{k}^i)r_{i}$, where $E$ is an encoding function; $E(w_{1}^i),\ldots, E(w_{k}^i)$ are the codewords of numbers $w_{1}^i,\ldots,w_{k}^i; s_{i}$ is the beginning of the text $u_{i}$ that does not contain a whole codeword; and  $r_{i}$ is the remainder of text  $u_{i}$ that does not contain a whole codeword.

As easy to see, the values  $w_{1}^i,\ldots,w_{k}^i$  as well as the remainder $r_{i}$ can be unambiguously determined by $u_{i}$ and the remainder $r_{i-1}$ of the previous portion of bytes.  Thus, we consider $u_{i}$ and $r_{i-1}$ as indices of predefined arrays $W_{1}, W_{2},\ldots,W_{k}, R$ containing the corresponding decoded numbers and a remainder,

$W_{1}[r_{i-1},u_{i}]= w_{1}^i,\ldots, W_{k}[r_{i-1},u_{i}]=w_{k}^i, R[r_{i-1},u_{i}]=r_{i}$.

 We get decoded numbers directly from these arrays.

Note that the concatenation $r_{i-1}s_{i}$ is also a codeword, if it is not empty. Some bits from the beginning of the number $E^{-1}(r_{i-1}s_{i})$ may be unambiguously obtained at the $(i-1)$-th iteration while others are obtained at the \emph{i}-th iteration. Thus, we can make correction assuming that $w_{1}^i$ and $w_{k}^i$ could be not the fully decoded numbers, but also the ending or the beginning of the decoded number binary representation respectfully. Values $w_{1}^i,\ldots,w_{k}^i$ corrected in this way we denote by $w_{1},\ldots,w_{k}$, eliminating the index $i$ for simplicity. Therefore, by $r_{i}$ we denote the ending of the text $u_{i}$, which cannot be decoded unambiguously at the \emph{i}-th iteration. Also, note that there is no need to store the first bit of numbers $w_{1},\ldots,w_{k}$, because it is always equal to one.

To illustrate how the method works, we apply this general byte aligned algorithm for the code $D_{2}$, assuming that at each iteration one byte is processed. The arrays $W_{1},...W_{k}$ are stored in the predefined table. Some rows of this table are shown in Table \ref{tab3}. The shortest codeword of $D_{2}$ has the form $110$. This implies that with little exception one byte can encompass no more than three full or partial codewords from  $D_{2}$.  The only option when the byte can cover four codewords fully or partially is the case  $0 110 110 x$, where $x$ is the last bit of the byte and the first bit of the fourth codeword. This bit can be attributed to the unprocessed remainder $r,$ and thus it is enough to store three resultant numbers.

Together with the numbers $w_1$, $w_2$, $w_3$ and the remainder $r$ we store the following values in each row of the table:
$|w_{i}|$ is the length of the \emph{i}-th number in bits (excluding the first bit);
$f_{i}$  is the  flag signaling if the codeword $w_{i}$ is the last in the current byte $(f_{i}=0)$ or not $(f_{i}=1)$.

Under the heading of Table \ref{tab3} there are rows written from top to bottom, which are used to decode the coded text $11000111 \; 01101011\; 11001011 \; 11101101 \; 10011000$.

\begin{table*}[!t]
\caption{Decoding table for bytewise method for the code $D_{2}$}
\label{tab3}
\centering
\begin{tabular}{|c|c||c|c|c|c|c|c|c|c|c|c|}
\hline
$r_{i-1}$&$u$&$w_1$&$|w_1|$&$f_1$&$w_2$&$|w_2|$&$f_2$&$w_3$&$|w_3|$&$f_3$&$r_i$\\ \hline
&11000111&&0&1&0011&4&0&&&&1\\ \hline
1&01101011&&0&1&1&1&0&&&&011\\ \hline
011&11001011&0111001&7&0&&&&&&&011\\ \hline
011&11101101&01111&5&1&&0&0&&&&1\\ \hline
1&10011000&&0&1&0&1&1&00&2&0&\\ \hline
\end{tabular}
\end{table*}

The structure of the second byte is shown in Fig. \ref{fig1}.

\begin{figure}[!t]
\centering
\begin{tabular}{|c|c|c|c|c|c|c|c|c|c|c|}
&$r_{i-1}$&\multicolumn{8}{|c|}{$i$-th byte}\\ \hline
$\ldots$&1&0&1&1&0&1&0&1&1\\ \hline
&\multicolumn{5}{|c|}{$E(w_1)$}&$E(w_2)$&\multicolumn{3}{|c|}{$r_i$}\\
\end{tabular}
\caption{Parsing of the byte 01101011}
\label{fig1}
\end{figure}

Let us examine the set of possible values of the remainder $r$. First, let us make the following comments:

\begin{enumerate}
  \item If some $(\Delta ,k)$-group is a part of the  byte composition, then it can be unambiguously decoded regardless of the next byte content, and, therefore, its bits will not be included in $r$.

\item If the byte ends with  $p\geq 3$ consecutive ones, then they will be decoded as $p-1$ ones regardless of the next byte content. In this case, the string $r$ consists of the last $1$, which during the decoding of the next byte will serve as an indication that the previous byte did not end with zero.

\item The string $10$ can be located only at the end or at the beginning of some $(\Delta ,k)$-group. In both cases, it can be decoded regardless of the next byte content: in the first case it is decoded together with the $(\Delta ,k)$-group, in which it is included. In the second case, it is decoded as $10$.
\end{enumerate}

It follows from the first of these observations that the sequence $r$ can not contain two consecutive zeros because such a situation is possible only if two zeros constitute a full $(\Delta ,k)$-group (then $r$ does not contain its bits), or when the first "0" is the end of one $(\Delta ,k)$-group, and the second $"0"$ is the beginning of the next group (in this case $r$ contains only the second zero). It follows from the second and third observations that the sequence $r$ can not contain three consecutive ones and the string $10$. Thus, we obtain a total 6 possible values of $r$: empty string, $0,1, 01, 11, 011$.

\begin{figure*}[!t]
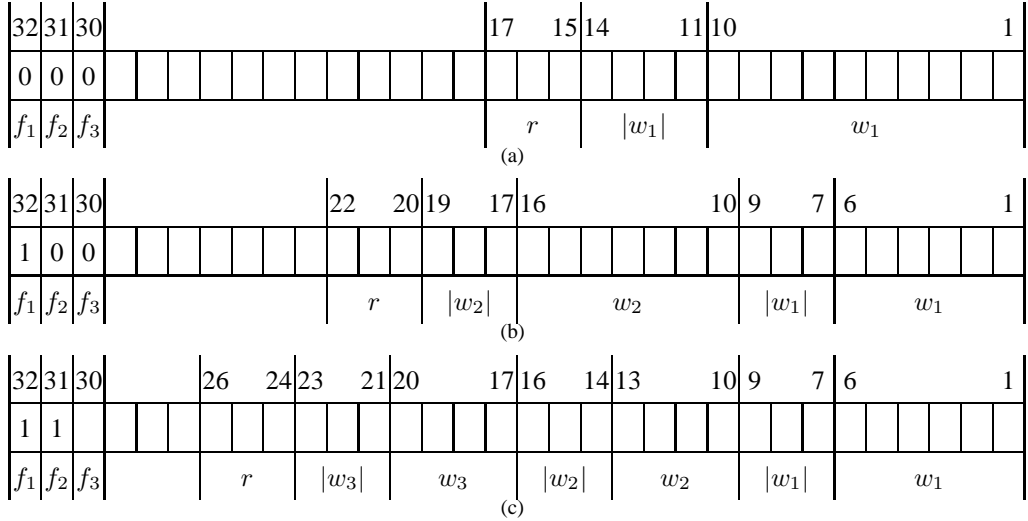

\setlength{\tabcolsep}{1pt}
\centering
\subfigure[\label{fig3a}]{
\begin{tabular}{|c|c|c|c|c|c|
c|c|c|c|c|c|c|c|c|c|c|c|c|c|c|c|c|c|c|c|c|c|c|c|c|c|}
32&31&30&\multicolumn{12}{|c|}{}&\multicolumn{1}{c}{17}&\multicolumn{1}{c}{}&\multicolumn{1}{c|}{15}&
\multicolumn{1}{c}{14}&\multicolumn{2}{c}{}&\multicolumn{1}{c|}{11}&
\multicolumn{1}{c}{10}&\multicolumn{8}{c}{}&\multicolumn{1}{c|}{1}\\ \hline
0&0&0&\hspace*{10pt}&\hspace*{10pt}&\hspace*{10pt}&\hspace*{10pt}&\hspace*{10pt}&\hspace*{10pt}&
\hspace*{10pt}&\hspace*{10pt}&\hspace*{10pt}&\hspace*{10pt}&\hspace*{10pt}&\hspace*{10pt}&\hspace*{10pt}&\hspace*{10pt}
&&\hspace*{10pt}&\hspace*{10pt}&\hspace*{10pt}&\hspace*{10pt}&\hspace*{10pt}&\hspace*{10pt}&
\hspace*{10pt}&\hspace*{10pt}&\hspace*{10pt}&\hspace*{10pt}&\hspace*{10pt}&\hspace*{10pt}&\hspace*{10pt}&\hspace*{10pt}\\ \hline
$f_1$&$f_2$&$f_3$&\multicolumn{12}{|c|}{}&\multicolumn{3}{|c|}{$r$}&
\multicolumn{4}{|c|}{$|w_1|$}&\multicolumn{10}{|c|}{$w_1$}\\
\end{tabular}
}
\subfigure[\label{fig3b}]{
\begin{tabular}{|c|c|c|c|c|c|
c|c|c|c|c|c|c|c|c|c|c|c|c|c|c|c|c|c|c|c|c|c|c|c|c|c|}
32&31&30&\multicolumn{7}{|c|}{}&\multicolumn{1}{c}{22}&\multicolumn{1}{c}{}&\multicolumn{1}{c|}{20}&
\multicolumn{1}{c}{19}&\multicolumn{1}{c}{}&\multicolumn{1}{c|}{17}&
\multicolumn{1}{c}{16}&\multicolumn{5}{c}{}&\multicolumn{1}{c|}{10}&
\multicolumn{1}{c}{9}&\multicolumn{1}{c}{}&\multicolumn{1}{c|}{7}&
\multicolumn{1}{c}{6}&\multicolumn{4}{c}{}&\multicolumn{1}{c|}{1}\\ \hline
1&0&0&\hspace*{10pt}&\hspace*{10pt}&\hspace*{10pt}&\hspace*{10pt}&\hspace*{10pt}&\hspace*{10pt}&
\hspace*{10pt}&\hspace*{10pt}&\hspace*{10pt}&\hspace*{10pt}&\hspace*{10pt}&\hspace*{10pt}&\hspace*{10pt}&\hspace*{10pt}
&\hspace*{10pt}&\hspace*{10pt}&\hspace*{10pt}&\hspace*{10pt}&\hspace*{10pt}&\hspace*{10pt}&\hspace*{10pt}&
\hspace*{10pt}&\hspace*{10pt}&\hspace*{10pt}&\hspace*{10pt}&\hspace*{10pt}&\hspace*{10pt}&\hspace*{10pt}&\hspace*{10pt}\\ \hline
$f_1$&$f_2$&$f_3$&\multicolumn{7}{|c|}{}&\multicolumn{3}{|c|}{$r$}&
\multicolumn{3}{|c|}{$|w_2|$}&\multicolumn{7}{|c|}{$w_2$}&
\multicolumn{3}{|c|}{$|w_1|$}&\multicolumn{6}{|c|}{$w_1$}\\
\end{tabular}
}
\subfigure[\label{fig3c}]{
\begin{tabular}{|c|c|c|c|c|c|
c|c|c|c|c|c|c|c|c|c|c|c|c|c|c|c|c|c|c|c|c|c|c|c|c|c|}
32&31&30&
\multicolumn{3}{|c|}{}&\multicolumn{1}{c}{26}&\multicolumn{1}{c}{}&\multicolumn{1}{c|}{24}&
\multicolumn{1}{c}{23}&\multicolumn{1}{c}{}&\multicolumn{1}{c|}{21}&
\multicolumn{1}{c}{20}&\multicolumn{2}{c}{}&\multicolumn{1}{c|}{17}&
\multicolumn{1}{c}{16}&\multicolumn{1}{c}{}&\multicolumn{1}{c|}{14}&
\multicolumn{1}{c}{13}&\multicolumn{2}{c}{}&\multicolumn{1}{c|}{10}&
\multicolumn{1}{c}{9}&\multicolumn{1}{c}{}&\multicolumn{1}{c|}{7}&
\multicolumn{1}{c}{6}&\multicolumn{4}{c}{}&\multicolumn{1}{c|}{1}\\ \hline
1&1&&\hspace*{10pt}&\hspace*{10pt}&\hspace*{10pt}&\hspace*{10pt}&\hspace*{10pt}&\hspace*{10pt}&
\hspace*{10pt}&\hspace*{10pt}&\hspace*{10pt}&\hspace*{10pt}&\hspace*{10pt}&\hspace*{10pt}&\hspace*{10pt}&\hspace*{10pt}
&\hspace*{10pt}&\hspace*{10pt}&\hspace*{10pt}&\hspace*{10pt}&\hspace*{10pt}&\hspace*{10pt}&\hspace*{10pt}&
\hspace*{10pt}&\hspace*{10pt}&\hspace*{10pt}&\hspace*{10pt}&\hspace*{10pt}&\hspace*{10pt}&\hspace*{10pt}&\hspace*{10pt}\\ \hline
$f_1$&$f_2$&$f_3$&\multicolumn{3}{|c|}{}&\multicolumn{3}{|c|}{$r$}&
\multicolumn{3}{|c|}{$|w_3|$}&\multicolumn{4}{|c|}{$w_3$}&
\multicolumn{3}{|c|}{$|w_2|$}&\multicolumn{4}{|c|}{$w_2$}&
\multicolumn{3}{|c|}{$|w_1|$}&\multicolumn{6}{|c|}{$w_1$}\\
\end{tabular}
}
\centering
\caption{Packing a string of decoding table into four-byte computer word}
\label{fig3}
\end{figure*}

Now we show that any row in Table \ref{tab3} can be "packed" into a single 32-bit machine word. We enumerate all possible values of $r$ by binary numbers from 0 to 5, and thus three bits are enough to store any such value. Note that if a certain flag $f_{i}$ is zero (this means that the word $w_{i}$ is not fully decoded), then there is no need to consider words $w_{i+1}, w_{i+2},...$, as well as flags $f_{i+1}, f_{i+2},...$,  as the code $w_{i}$  extends to the beginning of the string $r$ or to the right boundary of the byte. Denoting these values $f_{i},$ which can be disregarded, by zeroes, we obtain the following possible combinations of flag values $f_{1},f_{2},f_{3}: 000, 100$ and $11x$, where $x$-is an arbitrary binary value. For each of these cases we describe the special method of packing a row of Table \ref{tab3} into a four-byte word (Fig. \ref{fig3}). However, in any case we write the values $f_{1},f_{2},f_{3}$ into three most significant bits, the values $w_{1}, |w_{1}|$; $w_{2}, |w_{2}|$ (if available); $w_{3}, |w_{3}|$ (if available) and $r$, from the least significant to the most significant bits, in the specified order.

\textbf{$(f_{1}, f_{2}, f_{3})=000$.} In this case, the value $w_1$ takes no more than 10 bits. Indeed, consider first the case when $r_{i-1}=011$. If $f_{1}=0$, then the most significant bit of the byte $u_{i}$ can not be zero, since otherwise there would be a sequence $0110$, which means the end of the codeword and $f_{1}=1$. Assume, that all the bits of $u_{i}$ are ones. Then the last bit refers to $r_{i}$, and the length of the decoded value $w_{1}$ is $3 + 7 = 10$ bits. If $u_{i}$ contains the zero bit, then during decoding of $w_{i}$  the sequence of the form $01...10$ with more than 2 ones will be processed, which will correspond to one bit shorter piece of the  code $w_{i}.$ Therefore, the total bit length of $w_{i}$  will not exceed $3+8-1=10$ bits. If the value $r_{i-1}$ contains less than three bits, then the length $w_{i}$  obviously, cannot be longer than $8+2=10$ bits. Thus, in the case of $(f_{1}, f_{2}, f_{3})=000$, four bits are enough to store the value $|w_{1}|$, and, in general, the packing of a string of the Tab. 3 in a four-byte word appears as in Fig. \ref{fig3a}.

\textbf{$(f_{1}, f_{2}, f_{3})=100$.} In this case, the string concatenation $r_{i-1}u_{i}$ must contain the delimiter $0110$ or starts inside the delimiter. The value $w_{1}$ will be the longest if the delimiter is shifted to the right boundary of the byte. As the delimiter is not taken into consideration during  decoding, the value  $w_{1}$ will be obtained as a result of decoding at most 7 bits, and for reasons set out in the case $(f_{1}, f_{2}, f_{3})=000$,  the greatest possible length of $w_{1}$ will be one bit less, i.e. $|w_{1}|\leq 6$ and to store the value $|w_{1}|$ 3 bits are enough.

In the case $(f_{1}, f_{2}, f_{3})=100$ we also must store the value  $w_{2}$. Since the code  $w_{1}$ takes at least one bit of the byte  $u_{i}$, for the code  $w_{2}$ there remain no more than 7 bits, which requires 3 bits for the value $|w_{2}|$ and results in the packing as in Fig. \ref{fig3b}.

\textbf{$(f_{1}, f_{2}, f_{3})=11x$.} In this case, the code $w_{1}$ satisfies the same restrictions as in the case $(f_{1}, f_{2}, f_{3})=100$. The code $w_{2}$, which total length does not exceed 7 bits, must also contain a delimiter with no less than three bits. Thus, four bits are enough for value $w_{2}$, three bits for $|w_{2}|$. Since the code $w_{1}$ occupies at least one bit of the byte $u$, and the shortest code $w_{2}$ is $110$, then the length of encoded and decoded values $w_{3}$ is not longer than four bits. Thus, we get the packing shown in Fig. \ref{fig3c}.

Now we describe in detail the byte aligned algorithm of decoding for the code $D_{2}$ (Fig. \ref{program}). By $x << c$ denote the operation of shifting the value $x$ to the left and by $x >> c$ shifting to the right in $c$ bits (shift is not cyclic and new bits are filled with zeros).

The symbol $\&$ denotes the bitwise operation "and", and the symbol $|$ stands for the bitwise "or".  By $text_{i}$ we denote another byte of encoded text, by $t$ denote a string from Table \ref{tab3} packed in four-byte word. In the variable $w$ a decoded number is formed as the string concatenation $w_{1}, w_{2}$ or $w_{3}$, and in a variable \emph{len} the lengths of these strings are stored. The initial value $w$ consists of one "1" bit, then it shifts to the left, and the right bits are replaced by values $w_{1}, w_{2}$ or $w_{3}$ (from the relevant parts of the word $t$), and thus the most significant bit of $w$ always remains 1.

\begin{figure}[H]
\begin{tabbing}
\hspace{10mm}\=\hspace{10mm}\=\hspace{10mm}\=\hspace{10mm}\=\hspace{60mm}\=\\
$i\leftarrow 1;$\>\>\>\>\>//byte number of the encoded text\\
$r\leftarrow 0;$\\
$w\leftarrow 1;$\\
while (the end of the text is not reached) \{ \\
          \>$t \leftarrow TAB[r][text_i];$\>\>\>\>// read out 4-byte string in Tab. 3\\
          \>if($t\&\mathrm{0x80000000}$)   \{\>\>\>\>// if $f_1 = 1$\\
              \>\>$len \leftarrow (t >> 6)\&\mathrm{0x7};$\>\>\>// $len \leftarrow |w_1|$\\
              \>\>output $(w << len)|(t\&\mathrm{0x3F});$\>\>\>// decoded number: $w$ with \\
              \>\>\>\>\>// appended to the right 6 least significant bits of $t$\\
              \>\>$w \leftarrow 1;$\\
              \>\>if($x\&\mathrm{0x40000000}$)   \{\>\>\>// if $f_2 = 1$\\
                  \>\>\>$len \leftarrow (t >> 13)\&\mathrm{0x7};$\>\>// $len \leftarrow |w_2|$\\
                  \>\>\>output $(w << len)|((t >> 9)\&\mathrm{0xF});$\>\>// decoded number: $1w_2$\\
                  \>\>\>$w \leftarrow 1;$\\
                  \>\>\>$len \leftarrow (t >> 20)\&\mathrm{0x7};$\>\>// $len \leftarrow |w_3|$\\
                  \>\>\>if($t\&\mathrm{0x20000000}$)   \{\>\>// if $f_3 = 1$\\
                        \>\>\>\>output $(w << len)|((t >> 16)\&\mathrm{0xF});$\>// decoded number: $1w_3$\\
                        \>\>\>\>$w \leftarrow 1;$\\
                  \>\>\>\}   else\>\>// $(f_1, f_2, f_3)=110$\\
                        \>\>\>\>$w \leftarrow (w << len)|((t >> 16)\&\mathrm{0xF});$\>// $w \leftarrow 1w_3$\\
                  \>\>\>$r \leftarrow (t >> 23)\&\mathrm{0x7};$\>\>// $r$ in bits 24-26\\
              \>\>\} else   \{\>\>\>// $(f_1, f_2)=10$\\
                       \>\>\>$len \leftarrow (t >> 16)\&\mathrm{0x7};$ \>\>// $len \leftarrow |w_2|$\\
                       \>\>\>$w \leftarrow (w << len)|((t >> 9)\&\mathrm{0x7F});$ \>\>// $w \leftarrow 1w_3$\\
                       \>\>\>$r \leftarrow (t >> 19)\&7;$ \>\>// $r$ in bits 20-22\\
              \>\>\}\\
          \>\}   else   \{\>\>\>\>// if $f_1 = 0$\\
                   \>\>$len \leftarrow (t >> 10)\&\mathrm{0xF};$\>\>\>// $len \leftarrow |w_1|$\\
                   \>\>$w \leftarrow (w << len)|(t\&\mathrm{0x3FF});$ \>\>\>// append $w_1$ to  $w$ \\
                   \>\>$r \leftarrow (t >> 14)\&\mathrm{0x7};$ \>\>\>// $r$ in bits 15-17\\
          \>\}\\
          \>$i \leftarrow i+1;$\>\>\>\>// proceed to the next byte\\
\}
\end{tabbing}
\centering
\caption{Bytewise decoding algorithm for the code $D_2$}
\label{program}
\end{figure}

Let us estimate storage consumption of the method described above. For each of 6 possible values $r_{i-1}$ there exist $256$ values $u_{i}$, thus Table \ref{tab3} contains $6\times 256$ strings; 4 bytes are required to store each of them. Thus, the memory storage of the bytewise decoding method is 6 Kb.

Let us compare the space complexity of a given method with fast byte aligned methods used for decoding Fibonacci codes.
The most detailed study of them is presented in \cite{Kl1}, where three such methods are described.
The fastest of them is the method that involves using the table named Fib3. Its memory storage requires 21.4 Kb, i.e. more than 3.5 times greater than the method we propose.

 Time complexities of these methods were compared by numerical experiments. The random 20 million words fragment from English Wikipedia text corpus was encoded by the codes $D_{2}$ and Fib3 and then decoded by byte
aligned methods mentioned above. Time of decoding was measured. The experiment was repeated $100$ times, and the results were averaged. These results are shown in Table \ref{tab6}. As is seen, decoding of $D_{2}$ is about $20\%$ faster than that of Fib3. This mainly is due to the fact that the decoding of $D_2$ requires only one memory read operation at each iteration, after which all the other operations can be performed in processor registers very rapidly, while the mentioned above Fib3 decoding method requires 2 or 3 readings from one- or two-dimensional arrays at each iteration.

\begin{table}[!t]
\caption{Comparison of bytewise decoding methods complexity for codes $D_{2}$ and Fib3}
\label{tab6}
\centering
\begin{tabular}{|c|c|c|}
\hline
&Bytewise decoding of $D_2$&Bytewise decoding of Fib3\\ \hline
Memory&6K&21.4K \\ \hline
Time&$0.255s$&$0.321s$\\ \hline
\end{tabular}
\end{table}

\section{Compressing data by multi-delimiter codes}
Applicability of a code for information compressing is largely related to its density, which is measured by the number of codewords of the length not exceeding $n$. Let us first calculate the asymptotic density of the code $D_{2}$.
By $f_{n}$ denote the number of  codewords  in $D_{2}$ of the length $n$.

\begin{lemma} The following equality holds
\begin{equation}
\label{q}
f_{n}=f_{n-1}+f_{n-2}+f_{n-3}+f_{n-6}
\end{equation}
\end{lemma}

\begin{IEEEproof}
Applying formula (4) to parameters of the code $D_{2}(t=1, m_{1}=2)$ and taking into account that $T_{n}-T_{n-1}=0$ for $n\geq 6$, we obtain the following recurrent relation that is true for $n\geq 6:$

\begin{equation}
\label{q}
f_{n}=f_{n-1}+2f_{n-2}-f_{n-3}+f_{n-5}
\end{equation}

By induction, we prove that for $n \geq 7$  equality (8) is equivalent to (9). It is necessary to prove the equality of right parts (8) and (9), which after reductions
 takes the form  $f_{n-2}-f_{n-3}+f_{n-5}= f_{n-3}+f_{n-6}$. This gives the equality

\begin{equation}
\label{q4eq}
f_{n-2}+f_{n-5}=2f_{n-3}+f_{n-6}
\end{equation}

For $n = 7$ this equality is easy to check directly. Suppose, it holds for some $n \geq 7$. Express $f_{n-1}$ by using formula (9): $f_{n-1} = f_{n-2} + 2f_{n-3}-f_{n-4} + f_{n-6}$. It gives  $2f_{n-3}+ f_{n-6}= f_{n-1} - f_{n-2} + f_{n-4}$. Substituting this expression to the right side of (10), we obtain equality $f_{n-1} + f_{n-4} = 2f_{n-2} + f_{n-5}$,  which  coincides with equality (10), if replace $n$ by $n+1$.
\end{IEEEproof}

By $s_{n}$  denote the number of  codewords, which lengths do not exceed $n$, $s_{n}=\sum_{i=1}^n f_{i}$. Taking into account that $f_{3}= f_{4}=1, f_{5}=2, f_{6}=3$  and, summing over all indices $n\geq 7$ both parts of formula (8), we obtain:
\begin{eqnarray}
\label{q11}
s_{n}= \sum_{i=3}^6 f_{i} +\sum_{i=7}^n f_{i}=\nonumber \\
7+\sum_{i=7}^n(f_{i-1} + f_{1-2} + f_{i-3} + f_{i-6})
\end{eqnarray}

Note that the following identities hold:
\begin{eqnarray*}
\sum_{i=7}^n f_{i-1} = \sum_{i=6}^{n-1}f_{i}=s_{n-1}-4;\nonumber \\
\sum_{i=7}^n f_{i-2} = \sum_{i=5}^{n-2}f_{i}=s_{n-2}-2;\nonumber
\end{eqnarray*}
\begin{eqnarray*}
\sum_{i=7}^n f_{i-3} = \sum_{i=4}^{n-3}f_{i}=s_{n-3}-1;\nonumber \\
\sum_{i=7}^n f_{i-6} = \sum_{i=1}^{n-6}f_{i}=s_{n-6}.
\end{eqnarray*}

Substituting these expressions into formula (\ref{q11}), we obtain:

\begin{equation}
\label{q12}
s_{n}=s_{n-1} + s_{n-2} + s_{n-3} + s_{n-6}
\end{equation}

Since $s_{2}=s_{1}=s_{0}=s_{-1}=\cdots=0$,
$s_{3}=1, s_{4}=2, s_{5}=4, s_{6}=7$,
 the equality (\ref{q12}) holds for $n\geq6$. Formula (\ref{q12}) allows us to find the generating function $G(z)$ for $s_{n}$:

\begin{eqnarray}
\label{q3}
G(z)=\sum_{n=0}^{\infty}s_{n}z^{n}= z^{3}+ 2z^{4} + 4z^{5} +\nonumber \\
+\sum_{n=6}^{\infty}s_{n}z^{n}= z^{3}+ 2z^{4} + 4z^{5} +\nonumber \\
+\sum_{n=6}^{\infty}(s_{n-1}+s_{n-2}+s_{n-3}+ s_{n-6})z^{n}
\end{eqnarray}

Take into account the following equalities:

\begin{eqnarray*}
\sum_{n=6}^{\infty}s_{n-1}z^{n}= z\sum_{n=6}^{\infty}s_{n-1}z^{n-1}=\\
z\sum_{n=5}^{\infty}s_{n}z^{n}=zG(z)-z^{4}- 2z^{5};\\
\\
\sum_{n=6}^{\infty}s_{n-2}z^{n}= z^{2}\sum_{n=6}^{\infty}s_{n-2}z^{n-2}=\\
z^{2}\sum_{n=4}^{\infty}s_{n}z^{n}=z^{2}G(z)-z^{5};\\
\\
\sum_{n=6}^{\infty}s_{n-3}z^{n}= z^{3}\sum_{n=6}^{\infty}s_{n-3}z^{n-3}=\\
z^{3}\sum_{n=3}^{\infty}s_{n}z^{n}=z^{3}G(z);\\
\\
\sum_{n=6}^{\infty}s_{n-6}z^{n}= z^{6}\sum_{n=6}^{\infty}s_{n-6}z^{n-6}=\\
z^{6}\sum_{n=0} ^{\infty}s_{n}z^{n}=z^{6}G(z).
\end{eqnarray*}

Substituting these equalities  into  formula (13) and solving the resulting equation with respect to $G (z)$, we obtain:

 \[
G (z)= \frac{z^{3}+z^{4}+z^{5}}{1-z-z^{2}-z^{3}-z^{6}}=\frac{z^{3}}{1-2z+z^{3}-z^{4}}\]

 Decompose $G(z)$ to the sum of prime fractions

\begin{eqnarray}
\label{q14}
G(z)=\frac{-0.3618+0.2982i}{z-0.809-0.9816i} +\nonumber \\
+\frac{-0.3618+0.2982i}{z-0.809+0.9816i}-\nonumber \\
-\frac{0.1888}{z+1.1537}-\frac{0.0876}{z-0.5357},
\end{eqnarray}

 where $i$ is the imaginary unit, $i=\sqrt{-1}$.

 As seen from (\ref{q3}), the coefficient $s_{n}$ equals to the \emph{n}-th term of the  Maclaurin series for the function $G(z)$. If we expand function $g(z)=\frac{1}{z-a}$ into the Maclaurin series, then the $n$-th term equals to $\frac{x^n}{n!}g^{(n)}(0)=\frac{(-1)^n!x^n}{n!(-a)^n}=\frac{x^n}{a^n}$. Thus, the order of growth of $s_n$ is  determined by the value $1/a^n$, where the  value $a$ should be selected by the condition that $|a|$ is the smallest value among all fractions of the form $\frac{b}{z-a}$ in  formula (\ref{q14}).  This is the last fraction in (\ref{q14}). Thus, $a=0.5357$ and the order of growth of  $s_{n}$   is given by the expression

\begin{equation}
\label{q15}
\left(\frac{1}{0.5357}\right)^{n}\approx 1.867^{n}
\end{equation}

As shown in \cite{Kl1}, among the family of Fibonacci codes of higher orders the code Fib3 gives the best compression rate in the case of encoding natural language texts. The asymptotic density of this code is $1.839^{n}$. Thus, the code  $D_{2}$  is asymptotically denser than Fib3. It is also evident from the simple fact that the number of words of the length $n$ in the code $D_{2}$ determined by formula (8): $f_{n}=f_{n-1}+f_{n-2}+f_{n-3}+f_{n-6}$, while for the code Fib3 it is $f_{n}=f_{n-1}+f_{n-2}+f_{n-3}$.

Using the standard technique of generating functions, it is not difficult to calculate the asymptotic density of other multi-delimiter codes. For several such codes that may be of interest from the practical point of view, as well as for several Fibonacci codes, these values together with numbers of short codewords are given in Table \ref{tab7}.

\begin{table*}[!t]
\caption{The number of codewords of length $\leq n$ for some codes}
\label{tab7}
\centering
\begin{tabular}{|c|c|c|c|c|c|c|c|c|c|}
\hline
Code&Asymptotic&$n=2$&$n=3$&$n=4$&$n=5$&$n=6$&$n=7$&$n=8$&$n=15$\\
\hline
\multicolumn{10}{|c|}{The codes with the shortest codeword of the length 2}\\ \hline
Fib2&$1.618^n$&1&2&4&7&12&20&33&986\\ \hline
$D_1$&$1.755^n$&1&2&3&5&9&16&28&1432\\ \hline
$D_{1,2}$&$1.618^n$&1&3&5&7&10&16&27&799\\ \hline
$D_{1,3}$&$1.674^n$&1&2&4&7&11&18&30&1106\\ \hline
\multicolumn{10}{|c|}{The codes with the shortest codeword of the length 3}\\ \hline
Fib3&$1.839^n$&0&1&2&4&8&15&28&2031\\ \hline
$D_2$&$1.867^n$&0&1&2&4&7&13&24&1906\\ \hline
$D_{2,3}$&$1.785^n$&0&1&3&6&11&19&33&1874\\ \hline
$D_{2,4}$&$1.823^n$&0&1&2&5&9&17&30&1998\\ \hline
$D_{2,5}$&$1.844^n$&0&1&2&4&8&15&28&1999\\ \hline
$D_{2,3,4}$&$1.731^n$&0&1&3&7&13&23&39&1721\\ \hline
$D_{2,3,5}$&$1.755^n$&0&1&3&6&12&21&37&1833\\ \hline
$D_{2,4,5}$&$1.796^n$&0&1&2&5&10&19&34&2019\\ \hline
$D_{2,4,6}$&$1.809^n$&0&1&2&5&9&18&32&2032\\ \hline
\multicolumn{10}{|c|}{The codes with the shortest codeword of the length 4}\\ \hline
Fib4&$1.928^n$&0&0&1&2&4&8&16&1606\\ \hline
$D_3$&$1.933^n$&0&0&1&2&4&8&15&1510\\ \hline
\end{tabular}
\end{table*}

As seen, many multi-delimiter codes contain a larger number of short codewords than the comparing  Fibonacci codes with the same length of the shortest codeword. The "champions" are the codes $D_{2,3},$ $D_{2,3,4},$ $D_{2,3,5,}$ and $D_{2,4,5}$. They are the candidates for efficient compression. However, the code $D_{2,3,4}$ has quite low asymptotic density, which narrows its application area to only small alphabets. We investigate more thoroughly the other three codes together with the code  $D_2$, which has the highest asymptotic density in the class of codes with the shortest word of the length 3.

Compression efficiency of multi-delimiter codes was experimentally measured on different sources of English texts.
 Namely, we took the Bible (King James version), three other famous pieces of writing, and the full content of English Wikipedia.  The results are presented in Table \ref{tab8} in  terms of the average codeword length. We compared the performance of multi-delimiter codes and the Fibonacci code Fib3, which is taken as the base for comparisons. This code is known as the most efficient for natural language text compression among all Fibonacci codes.

\begin{table*}[!t]
\caption{Empirical comparison of compression rate (the average codeword length) of Fib3 and some multi-delimiter codes}
\label{tab8}
\centering
\begin{tabular}
{|l|l|l|l|l|l|l|}
\hline
Source & Alphabet size & Fib3 & $D_{2}$ & $D_{2,3}$ & $D_{2,3,5}$ &$D_{2,4,5}$  \\
\hline

\hline
Bible KJV  &12,452 &$9.21$  &$9.35 (+1.6\%)$ &$9.03 (-2\%)$ & $8.95(-2.8\%)$ & $9.04(-1.8\%)$ \\
\hline

\hline
Hamlet, Shakespeare & 4,501 &$10.0$  & $10.16 (+1.6\%)$ &$9.82 (-1.8\%)$  &$9.74(-2.5\%)$ & $9.81(-1.9\%)$   \\
\hline

\hline
Robinson Crusoe, D. Defoe & 5,994 & $9.4$ &$9.55 (+1,6\%)$ & $9.19 (-2.2\%)$ & $9.12 (-3\%)$ & $9.21 (-2\%)$   \\
\hline

\hline
Oliver Twist, C. Dickens & 10,027 & $10.06$ &$10.21 (+1,5\%)$ &$9.91 (-1.6\%)$ &$9.84 (-2.3\%)$ & $9.89 (-1.7\%)$   \\
\hline

\hline
English Wikipedia & 5,487,696 &$11.585$  &$11.696 (+1\%)$ &$11.521 (-0.6\%)$  &$11.517 (-0.6\%)$ & $11.497 (-0.8\%)$     \\
\hline

\end{tabular}
\end{table*}

As seen, the codes with 2 and 3 delimiters outperform the Fib3 code. For example, the average codeword length for the  code $D_{2,3,5}$ is about $2-3\%$ less than that  for the code Fib3, if the alphabet size is around 10K words. This is a significant difference if we take into account that the code Fib3  exceeds the entropy bound only by $5-6\%$ for English texts, as reported in \cite{Kl1}. Since the asymptotic density of multi-delimiter codes is lower, their overheads over Fib3 decreases as alphabet size grows. However, codes with 2 and 3 delimiters are still superior even for Wikipedia, which is one of the largest known natural language text corpus up to date, containing over 5 million different words.

The code Fib3, in comparison with the multi-delimiter codes, also has a drawback, which refers to the characteristic of the instantaneous separation that is important for searching a word in the compressed file without its decompression. As Fib3, so multi-delimiter codes as well as  other codes used for text compression are characterized by the following: if a certain bit sequence $w$ occurs in a compressed file, we can not guarantee that it truly corresponds to the occurrence of the whole codeword $w$, since it could be the suffix of another codeword. In multi-delimiter codes to check if $w$ is truly a separate codeword it is enough to consider a fixed number of bits that precede $w$. For example, it is enough to check 4 bits for the code $D_2$. If they turn out to be $0110$, then $w$ is a codeword, otherwise it is not. However, it is not enough to check any fixed number of bits preceding a codeword in the code Fib3, since a delimiter and the shortest word in this code is 111.  Several such codewords can "stick together" if they are adjacent. As one of the ways to avoid this problem, in \cite{Kl1} it is proposed to extract the shortest codeword $111$ from the code Fib3. However, the density and compression efficiency of the code obtained in this way is significantly worse than those for all the codes discussed above, including $D_{2}$.

\section {Conclusion}

In this paper we introduce a new family of splittable codes that are based on encoding sequences of ordered integer pairs. Splittable codes form a rich set of codes that include the $(2,3)$-codes,  the Fibonacci codes of higher orders and the multi-delimiter codes.

The multi-delimiter codes are of special interest. They possess all properties known for the Fibonacci codes such as completeness, universality, simple vocabulary representation, and strong robustness. But also they have some more advantages:

\begin{itemize}{\IEEEsetlabelwidth{(iii)}}
\item[(i)]	Adaptability.  Varying delimiters we can adapt a  multi-delimiter code to a given source probability distribution and an alphabet size.

\item[(ii)] Better compression rate for natural language text compressing.

\item[(iii)] Good computer performance minimizing time and storage overheads.

\item[(iv)] Instantaneous separation of codewords allowing faster compressed search.

\end{itemize}

The set of multi-delimiter codes together with the set of Fibonacci codes can be useful in many practical applications.

%

\begin{IEEEbiographynophoto}{Anatoly Anisimov} (M'12) was born on June 15, 1948 in South Sakhalin, Russia. He was awarded the diploma in mathematics (1970), PhD (1973) and Doctor's degree (1983) in Computer Science from Taras Shevchenko National University of Kyiv, Ukraine.

He is Professor of Computer Science at Taras Shevchenko National University of Kyiv since 1984; Head of the Department of Mathematical Informatics. In 1977 he worked as a Visiting Scientist at the Stanford University, Stanford, USA. He coauthored with D. Knuth the paper "Inhomogeneous Sorting", published in the International Journal of Computer and Information Sciences, vol. 8, no. 4, 1979. His research interests include algorithms, codes, parallel programming, information security, artificial intelligence.

Prof. Anisimov is a Corresponding Member of the National Academy of Sciences of Ukraine.

\end{IEEEbiographynophoto}

\begin{IEEEbiographynophoto}{Igor Zavadskyi} was born on May 29, 1974 in Kyiv, Ukraine. He was awarded the diploma in applied mathematics (1996) and PhD in Computer Science (2001) from Taras Shevchenko National University of Kyiv, Ukraine.

He is Associate Professor of Computer Science at Taras Shevchenko National University of Kyiv since 2007. His research interests include codes, parallel programming, optic and quantum computing, database management systems.

\end{IEEEbiographynophoto}

\end{document}